%% file: oprcp2.tex
\documentclass[twocolumn,longauthor]{aastex61}
\usepackage{amsmath}
\usepackage{amssymb}
\usepackage{graphicx}
\usepackage{wasysym}
\usepackage{natbib}
\usepackage{txfonts}

\newcommand{\xo}[2]{\overset{#1}{#2}} 
\newcommand{\mytitle}[0]{Cosmological tests of the gravastar hypothesis}
\newcommand{\myauthor}[0]{K.~A.~S.~Croker}
\newcommand{\rmd}[0]{\mathrm{d}}

\newcommand{\derp}[1]{\frac{\rmd #1}{\rmd t}}

\shorttitle{\mytitle}
\shortauthors{\myauthor}
\begin{document}
\title{\mytitle}
\author[0000-0002-6917-0214]{\myauthor}
\affil{Department of Physics \& Astronomy, University of Hawai`i at M\=anoa, 2505 Correa Rd., Honolulu, Hawai`i 96822, USA}
\email{kcroker@phys.hawaii.edu}

\date{September 3, 2017}

\begin{abstract}
Gravitational vacuum stars (gravastars) have become a viable
theoretical alternative to the black hole (BH) end-stage of stellar
evolution.  These objects gravitate in vacuum like BHs, yet have no
event horizons.  In this paper, we present tests of the gravastar
hypothesis within flat Friedmann cosmology.  Such tests are
complementary to optical and gravitational wave merger signatures,
which have uncertainties dominated by the poorly constrained gravastar
crust.  We motivate our analysis directly from the action principle,
and show that a population of gravastars must induce a time-dependent
dark energy (DE).  The possibility of such a contribution has been
overlooked due to a subtlety in the \emph{de facto} definition of the
isotropic and homogeneous stress.  Conservation of stress-energy
directly relates the time evolution of this DE contribution to the
measured BH comoving mass function and a gravastar population crust
parameter $\eta$.  We show that a population of gravastars formed
between redshift $8 \lesssim z \lesssim 20$ readily produces the present-day DE
density over a large range of $\eta$, providing a natural resolution
to the coincidence problem.  We compute an effective DE equation of
state $w_\mathrm{eff}(a)$ as a function of present-epoch BH population
observables and $\eta$.  Using a BH population model developed from
the cosmic star formation history, we obtain $w_0$ and $w_a$
consistent with Planck best-fit values.  In summary, the gravastar
hypothesis leads to an unexpected correlation between the BH
population and the magnitude and time-evolution of DE.
\end{abstract}


\section{Introduction}
The recent direct observation of gravitational radiation from
ultracompact, massive, object mergers provides definitive proof of the
existence of objects with exterior geometries consistent with the
classical black hole (BH) solutions.  Unfortunately, the various
boundaries and interiors of the classical BH geometries are severely
pathological.  In fact, the predicted physical curvature singularities
and closed timelike curves had already motivated the exploration of BH
``mimickers.'' Such solutions to classical GR appear as classical BH
geometries, as perceived by exterior vacuum observers, but replace the
interior with another GR solution.  \citet{gliner1966algebraic} is
usually recognized as the first researcher to suggest that ``$\mu$
vacuum,'' localized regions of vacuum with large cosmological
constant, could be relevant during gravitational collapse.

A simplified scenario, which motivates more physically plausible
constructions, is the following exact, spherically symmetric, GR
solution
\begin{align}
\rmd s^2 = -\beta \rmd t^2 + \beta^{-1}\rmd r^2 + r^2\left[\rmd \theta^2 + \sin^2\theta~\rmd\phi^2 \right] \label{eqn:gstar_skeleton}
\end{align}
with
\begin{align}
\beta(r) \equiv
\begin{cases}
1 - \left(r/2M\right)^2 & r \leqslant 2M \\
1 - 2M/r & r > 2M.
\end{cases}\label{eqn:gstar_skeleton_regions}
\end{align}
This solution is a Schwarzchild BH exterior for $r > 2M$, and a vacuum
with cosmological constant interior to $r < 2M$.  This solution is
static, consistent with the intuition that material under tension
stays bound.  Note that the interior energy density correctly
integrates to $M$.  This means that a \emph{vacuum} observer at
infinity will perceive a point mass $M$.  The interior region contains
no physical singularity, but the infinite pressure gradient at $r =
2M$ in Eqn.~(\ref{eqn:gstar_skeleton}) suggests consideration of
slightly more sophisticated GR solutions.

\citet[][Eqn.~(14)]{dymnikova1992vacuum} produced a ``G-lump'' model
without an infinite pressure gradient.  About a decade later,
\citet{mazur2004gravitational} and \citet{chapline2003quantum}
proposed additional realizations of what is now called a gravitational
vacuum star (gravastar).  The essential feature of viable gravastars
is a ``crust'' region of finite thickness slightly beyond $r = 2M$,
which surrounds the de Sitter core.  These objects have no trapped
surface: each interior is in causal contact with the outside universe.
In fact, this ability to resolve the BH information paradox sparked
renewed interest in BH mimickers during the ``firewall'' debates
initiated by \citet{almheiri2013black}.  Theoretical checks on
gravastar stability to perturbations \citep[e.g.][]{visser2004stable,
  lobo2006stable, debenedictis2006gravastar} and stability to rotation
\citep[e.g.][]{chirenti2008ergoregion, uchikata2015slowly,
  maggio2017exotic} have found large ranges of viable parameters
describing the crust.  Phenomenological checks from distinct merger
ringdown signatures, such as an altered quasinormal mode spectrum
\citep[e.g.][]{chirenti2007tell} have been computed.  In addition,
possible optical signatures from the crust itself
\citep[e.g.][]{broderick2007all}, accretion disks
\citep[e.g.][]{harko2009can}, and even direct lensing
\citep{sakai2014gravastar} have been investigated.

The first direct detection of gravitational radiation from massive
compact object mergers has greatly stimulated critical analysis of the
gravastar scenario.  \citet{bhagwat2016spectroscopic,
  bhagwat2017erratum} show that the existing aLIGO interferometer and
the proposed Cosmic Explorer and Einstein telescopes can resolve the
higher harmonics, which could probe the existence of gravastars.  On
the other hand, \citet{chirenti2016did} have already claimed, using
the ringdown, that GW150914 likely did not result in a gravastar final
state.  \citet{yunes2016theoretical} criticize the
\citet{chirenti2016did} claim as premature, yet argue that ringdown
decay faster than the light crossing time of the remnant poses the
more severe challenge.

In all phenomenology to date, however, gravastar signatures depend
critically on the unknown properties of the crust.  In this paper, we
address this problem and develop complementary observational
signatures dominated instead by the known properties of the
core. Before proceeding, we first comment on consistency with
Birkhoff's theorem, which gravitationally decouples the interior
properties of a localized object from its surroundings.  A necessary
condition for Birkhoff's theorem is vacuum boundary conditions.  Thus,
we will consider gravastars within Friedmann cosmology, which is
nowhere vacuum.

Consistent with the theoretical motivation to resolve the BH
information paradox, we will replace all BHs with gravastars.  The
following falsifiable experimental prediction for the physical dark energy (DE) density $\Omega_\Lambda^\mathrm{eff}$
\begin{align}
\kappa(a,a_c) &\equiv \frac{1}{a^3}\exp\left(3 \int_{a_c}^a \frac{\eta(a')}{a'} ~\rmd a'\right) \\
\Omega_\Lambda^\mathrm{eff}(a) &= \frac{1}{a^3\kappa(a)}\int_0^a\frac{\rmd \Delta_\mathrm{BH}}{\rmd a'} \kappa(a')~\rmd a'
\end{align}
will then be obtained.  Here $\eta$ is a small parameter which
measures the deviation of the gravastar contribution from pure de
Sitter due to crusts, $a_c$ is a cutoff before which there are no
gravastars, and $\Delta_\mathrm{BH}$ is the comoving BH density, which
has just begun to be constrained by aLIGO.  As detailed by
\citet{dwyer2015gravitational}, the proposed Cosmic Explorer
interferometer will be able to definitively constrain
$\Delta_\mathrm{BH}(a)$ with $\sim 10^{5}$ events per year, and out to
redshift $z \sim 10$.  The schematic gravastar given in
Eqn.~(\ref{eqn:gstar_skeleton}) has no crust, so $\eta \equiv 0$ and
the prediction becomes entirely parameter-free
\begin{align}
\Omega_\Lambda^\mathrm{eff}(a) &= \int_0^a\frac{\rmd \Delta_\mathrm{BH}}{\rmd a'}\frac{1}{a'^3}~\rmd a'.
\end{align}

The rest of this paper is organized as follows.  In
\S\ref{sec:stress}, we motivate a departure from the \emph{de facto}
Friedmann source (DFFS).  We then define notation that simplifies our
calculations and clarifies the physical observables.  In
\S\ref{sec:gravastars}, we construct the appropriate cosmological
source from an assumed population of gravastars.  In
\S\ref{sec:thenut}, we produce the fundamental quantitative prediction
relating the DE density to the observed BH population and the
gravastar crust parameter $\eta$.  In \S\ref{sec:implications}, we
proceed chronologically through the history of the Universe and
consider the influence of a gravastar population at three different
epochs.  In \S\ref{sec:primordial}, we consider the formation of
primordial gravastars and essentially exclude their production.  In
\S\ref{sec:darkages}, we demonstrate that stellar processes between
redshift $8 < z < 20$ can produce enough gravastars to account for the
present-day DE density.  In \S\ref{sec:planck_constraint}, we analyze
the gravastar-induced DE at late times ($z < 5$) in the dark fluid
framework described in \citet{ade2016planckXIV}.  Here, we produce
predictions for the DE equation of state parameters $w_0$ and $w_a$ in
terms of present-day BH population observables.

A comprehensive set of appendices presents material complementary to
the phenomenological results of this paper.
Appendix~\ref{sec:assumptions} highlights non-trivial assumptions
fundamental to the DFFS.  Appendix~\ref{sec:action} clarifies the
spatially-averaged nature of the Friedmann source by deriving
Friedmann's equations directly from the principle of stationary
action.  Appendix~\ref{sec:inside} establishes that the positive
pressure contributions of typical astrophysical systems can be
ignored.  Appendix~\ref{sec:causality} discusses how causality is
maintained, given a non-trivial time dependence in the background
cosmological source.  Appendix~\ref{sec:distribution} develops a
simple model for the BH population in terms of the stellar population.
Appendix~\ref{sec:dispersion} gives an upper bound for the
pressures of large, virialized systems like clusters.

Except within the appendices, we take the time unit as the reciprocal
present-day Hubble constant $H_0^{-1}$, the density unit as the
critical density today $\rho_\mathrm{cr}$, and set the speed of light
$c \equiv 1$.  For efficiency, we will often describe quantities with
scale factor or redshift dependence as time-dependent.  We will also
freely use either scale factor $a$ or redshift $z$, depending on which
produces the more compact quantitative expression.  Though we replace
all BHs with gravastars, for consistency with existing astrophysical
literature, we will often refer to these objects as BHs.

\section{Isotropic and homogeneous stress in Friedmann cosmology}
\label{sec:stress}
Given the symmetries of the RW metric, RW observers cannot distinguish
points: in a homogeneous and isotropic universe, every point is
observationally identical.  In other words, a RW universe does not
contain any notion of interior or exterior.  Perhaps surprisingly, the
\emph{de facto} Friedmann source (DFFS) implicitly violates
homogeneity and isotropy.  As detailed in
Appendix~\ref{sec:assumptions}, the DFFS assumes the ability to define
a notion of interior and exterior, distinct for every mass
distribution.  This assumption is used to reduce spatially extended
systems to effective point masses.  These point masses are then
considered to contribute only as a pressure-free ``dust.''

Consistency, however, requires that the source to Einstein's equations
exhibit the same symmetries as the metric.  This requirement holds at
each order in a perturbative expansion of Einstein's equations.  In
particular, the spatially uniform background source in Friedmann
cosmology cannot contain any implicit notion of interior or exterior.
In Appendix~\ref{sec:action}, we construct a consistent Friedmann
source (CFS) by application of the principle of stationary action to
an inhomogeneous fluid ansatz.  The resulting zero-order source simply
becomes a flat spatial average.

In contrast to the DFFS, the spatial average present within the CFS
necessarily incorporates \emph{all} pressures: cluster interiors,
stellar interiors and, should they exist, gravastar interiors.  In
Appendix~\ref{sec:inside}, we show that accounting for positive
astrophysical pressure contributions makes no observable changes.  The
effect of any \emph{negative} pressure contributions, however, can
critically influence the background expansion.  An averaged gravastar
interior contribution is both negative and strong, with
$|\mathcal{P}_s| \sim \rho_s$.  This feature of gravastars ultimately
leads to the unexpected and quantitative relation between the
expansion rate and the BH population.

\subsection{Two-component approximation}
Let barred variables represent physical, as opposed to comoving,
background quantities.  We investigate the CFS with two contributions.
The first contribution, the $s$-contribution, is an approximation to
the influence of gravitationally bound systems with pressure
\begin{align}
\xo{s}{T^\mu_\nu} &\equiv -\bar{\rho}_s(a) \delta^\mu_0\delta^0_\nu + \bar{\mathcal{P}}_s(a)\delta^\mu_i\delta^i_\nu \label{eqn:star_source_common}.
\end{align}
It is an approximation because we do not attempt to track all
pressures within the universe.  For the primary discussion of this
work, the $s$-contribution will be exclusively from gravastars.  For
the discussion in Appendix~\ref{sec:inside}, justifying why this is an
excellent approximation, the $s$-contribution will be a severe
upper-bound on positive pressure contributions from typical
astrophysical systems.  

As is well known, the constituents of tightly bound gravitational
systems decouple from the expansion.  The number density of these
systems then dilutes with the physical volume.  This behavior
motivates the following definitions
\begin{align}
\rho_s(a) &\equiv \bar{\rho}_s(a)a^3 \label{eqn:defn_rhos} \\
\mathcal{P}_s(a) &\equiv \bar{\mathcal{P}}_s(a)a^3
\end{align}
allowing us to to rewrite Eqn.~(\ref{eqn:star_source_common}) in an
entirely equivalent way
\begin{align}
\xo{s}{T^\mu_\nu} \equiv -\frac{\rho_s(a)}{a^3} \delta^\mu_0\delta^0_\nu + \frac{\mathcal{P}_s(a)}{a^3}\delta^\mu_i\delta^i_\nu. \label{eqn:star_source}
\end{align}
The explicit appearance of $a^3$ factors and time-dependent comoving
quantities greatly simplifies many computations and will help to
emphasize the difference between the DFFS and the CFS.  

The second contribution, the $0$-contribution, comes from collisionless
matter
\begin{align}
\xo{0}{T^\mu_\nu} &\equiv -\frac{\rho_0(a)}{a^3} \delta^\mu_0\delta^0_\nu \label{eqn:normal_source},
\end{align}
and is identical to the analogous term of the DFFS whenever $\rmd
\rho_0 / \rmd a = 0$.  This derivative statement is an example of the
utility of time-dependent comoving densities, when highlighting the
differences between the DFFS and the CFS.

In the gravastar scenario, effectively collisionless matter is
processed into localized regions of de Sitter space.  We thus populate
$\rho_s(a)$ by depleting an appropriate fraction $\Delta(a)$ of the
collisionless contribution
\begin{align}
\rho_0(a) \equiv \Omega_m[1 - \Delta(a)]. \label{eqn:rho0}
\end{align}
Here $\Omega_m$ is the observed matter fraction today.  The depletion
fraction $\Delta(a)$ is determined through astrophysical observation.

The CFS does not conserve particle number by construction, but this is
not required by Einstein's equations~\citep[\S2.10]{WeinbergGR}.
Covariant conservation of energy and momentum
\begin{align}
\nabla_\mu \left[\xo{0}{T^\mu_\nu} + \xo{s}{T^\mu_\nu}\right] = 0, \label{eqn:cons}
\end{align}
however, necessarily populates both $\rho_s(a)$ and $\mathcal{P}_s(a)$
in a manner consistent with Einstein's equations.  Thus,
time-dependence in comoving $\rho_0(a)$, $\rho_s(a)$, and
$\mathcal{P}_s(a)$ does not violate thermodynamics or energy
conservation.  For a discussion of causality in the context of a
non-trivially time-dependent background, we direct the reader to
Appendix~\ref{sec:causality}.

\section{Gravastar contribution to the Friedmann equations}
\label{sec:gravastars}
A dynamically stable gravastar, with physically plausible structure,
is more sophisticated than Eqns.~(\ref{eqn:gstar_skeleton}) and
(\ref{eqn:gstar_skeleton_regions}).  Both \citet{dymnikova1992vacuum}
and \citet{mazur2004gravitational} introduce a transition region, which
ultimately encloses a dark energy interior.  The possible nature of
this ``crust'' has been studied extensively by
\citet{visser2004stable, martin2012generic} and others.  There is
consensus that substantial freedom exists in the construction of
gravastar models.  This freedom, while constrained, permits a range of
crust thicknesses and equations of state.  We will consider the
following structure
\begin{itemize}
\item{$r \leqslant 2M$: de Sitter interior}
\item{$2M < r \leqslant 2M + \ell$: crust region}
\item{$r > 2M + \ell$: vacuum exterior}
\end{itemize}
where $M$ is a gravastar mass and $\ell$ is a crust thickness.  Note
that we have simplified the model by placing the inner radius at the
Schwarzchild radius.  In any physical gravastar, this inner radius
must be slightly beyond the Schwarzchild radius: this prevents the
formation of a trapped surface.  For our purposes, this nuance will
not matter.  The cosmological contribution from a plausible gravastar
population thus consists of two components
\begin{align}
\rho_s &\equiv \rho_\mathrm{int} + \rho_\mathrm{crust} \label{eqn:defn_rhos} \\
\mathcal{P}_s &\equiv -\rho_\mathrm{int} + \mathcal{P}_\mathrm{crust} \label{eqn:defn_Ps}.
\end{align}
Note that $\rho_\mathrm{int}$, $\rho_\mathrm{crust}$, and
$\mathcal{P}_\mathrm{crust}$ are not independent degrees of freedom
because each gravastar has a model-dependent internal structure.

We may define an equation of state for the gravastar source
\begin{align}
w_s &\equiv \frac{-\rho_\mathrm{int} + \mathcal{P}_\mathrm{crust}}{\rho_\mathrm{int} + \rho_\mathrm{crust}}.
\end{align}
\citet{mazur2004gravitational} assert that most of the energy density
is expected to reside within each core and that $\mathcal{P} \sim
\rho$ within each crust.  It is thus reasonable to expand the
aggregate quantities and discard higher-order terms 
\begin{align}
w_s &= -1 + \frac{\mathcal{P}_\mathrm{crust}}{\rho_\mathrm{int}}\left(1 - \frac{\rho_\mathrm{crust}}{\rho_\mathrm{int}}  + \dots \right)\\
&\simeq -1 + \frac{\mathcal{P}_\mathrm{crust}}{\rho_\mathrm{int}} \label{eqn:ws}.
\end{align}
This expression successfully isolates ignorance of the crust to a
small dimensionless parameter, which we now define as
\begin{align}
\eta(a) \equiv \frac{\mathcal{P}_\mathrm{crust}}{\rho_\mathrm{int}}.
\end{align}
To avoid confusion, we emphasize now that $w_s$ is not the Dark Energy
equation of state as constrained by Planck.  The relation between
these two quantities will be explored thoroughly in
\S\ref{sec:planck_constraint}.  

\subsection{Relation of $\rho_s$ to the BH population} 
\label{sec:thenut}
Let $\Delta_\mathrm{BH}$ be the comoving coordinate BH mass density.
The formation of gravastars, and any subsequent accretion, depletes
the baryon population.  This decreases $\rho_0$ by
$\Delta_\mathrm{BH}$ as per Eqn~(\ref{eqn:rho0}).  Consistent with the
results of Appendix~\ref{sec:inside}, to very good approximation, we
may regard all (non-gravastar) positive-pressure contributions as
collisionless within the Friedmann source.  Thus, the time-variation
of $\rho_0$ comes entirely from the time-variation of
$\Delta_\mathrm{BH}$.  The conservation Eqn.~(\ref{eqn:cons}) then
becomes
\begin{align}
\derp{\rho_s} + 3w_sH\rho_s = \derp{\Delta_\mathrm{BH}}. \label{eqn:cons_bh}
\end{align}
Switching from coordinate time $t$ to scale factor $a$ gives
\begin{align}
\frac{\rmd \rho_s}{\rmd a} + 3w_s\frac{\rho_s}{a} = \frac{\rmd \Delta_\mathrm{BH}}{\rmd a}.
\end{align}
This equation can be separated with an integrating factor 
\begin{align}
\kappa(a,a_c) &\equiv \exp\left(3\int_{a_c}^a \frac{w_s(a')}{a'}~\rmd a'\right) \label{eqn:kappa} \\
& = \frac{1}{a^3}\exp\left(3 \int_{a_c}^a \frac{\eta(a')}{a'} ~\rmd a'\right) 
\end{align}
where we have defined the proportionality to be unity and $a_c$ is a
cutoff below which there are no gravastars.  If we omit the cutoff
parameter for $\kappa$, then we implicitly assume $a_c$.  The resulting
energy density is
\begin{align}
\rho_s = \frac{1}{\kappa(a)}\int_0^a\frac{\rmd \Delta_\mathrm{BH}}{\rmd a'} \kappa(a')~\rmd a' \label{eqn:prediction},
\end{align}
where the lower-bound does not depend on $a_c$ because
$\Delta_\mathrm{BH}/\rmd a$ vanishes by definition before $a_c$.  When
convenient, we will study the phenomenology of gravastars with the
single parameter $\eta$ held fixed.  This is consistent with other
phenomenological studies of gravastars.  In practice, we will find
that time-variation of $\eta$ does not affect the essential
cosmological signatures of a gravastar population.

\subsubsection{Physical interpretation of DE} 
To develop some intuition for Eqn.~(\ref{eqn:prediction}), consider a
sequence of instantaneous conversions
\begin{align}
\frac{\rmd \Delta_\mathrm{BH}}{\rmd a} \equiv \sum_{n} Q_n \delta(a - a_n)
\end{align}
where $Q_n$ is the comoving density of baryons instantaneously
converted at $a_n$.  Substitution into Eqn.~(\ref{eqn:prediction}) gives 
\begin{align}
\rho_s = \sum_n^{a_n \leqslant a} Q_n \frac{\kappa(a_n)}{\kappa(a)}.
\end{align}
Explicitly, at the instant of the $m$-th conversion $a_m$, we have the
following comoving density of DE
\begin{align}
\rho_s(a_m) = Q_m + \sum_n^{a_n < a_m} Q_n \frac{\kappa(a_n)}{\kappa(a_m)}.
\end{align}
Thus, at the instant of conversion, $\rho_s$ agrees with a spatial
average over the newly formed gravastars' interiors.  In other words,
a quantity of baryonic mass has been converted to an equal quantity of
DE.  Beyond the instant of conversion, however, the contribution due
to gravastars dilutes more slowly than the physical volume expansion.
For example, in the de Sitter limit of $\eta \to 0$, we find that
$\kappa \to a^{-3}$ and
\begin{align}
\bar{\rho}_s(a) = \sum_n^{a_n \leqslant a} \frac{Q_n}{a_n^3} \qquad (\eta \equiv 0).
\end{align}
In this case, the physical baryon density becomes ``frozen in'' at the
time of conversion.  To maintain locality in general, one might
conclude that each gravastars' local mass $Q_n$ must increase such
that
\begin{align}
Q_n \propto \kappa a^3,
\end{align}
but this would violate the Strong Equivalence
Principle \citep[e.g.][]{TEGP}.  Though such behavior is conceivable
within scalar-tensor theories, we will remain focused on gravastars
within GR.  

Evidently, we cannot make any quantitative statements about the
spatial location of the subsequent energy averaged to produce
$\rho_s$.  This is by choice, as our analysis is designed to minimize
the impact of the unknown local details that produce the gravastar.
Since we have accomplished this through covariant conservation of
stress-energy, we can proceed with confidence that our predictions are
entirely consistent within GR.  Strictly speaking, however,
Eqn.~(\ref{eqn:ws}) must be interpreted as a motivation for $\eta$,
and nothing more.

\section{Observational implications}
\label{sec:implications}
To begin constraining the gravastar hypothesis, we determine when
gravastar production is observationally viable.  Our discussion will
proceed chronologically through the history of the Universe.  This
will inform our subsequent analysis: $\Omega_\Lambda^\mathrm{eff}$ at
late times will depend on an order-of-magnitude understanding of its
early-time behavior.

\subsection{Constraint on primordial pBH $\beta$ as a function of temperature}
\label{sec:primordial}
\begin{figure}
\caption{\label{fig:primordial}Maximum fraction $\beta$ of the
  critical density that can become primordial black holes as a
  function of gravastar crust parameter $\eta$.  The constraint is
  determined by requiring the population-induced DE not exceed
  $\Omega_\Lambda$.  Increasing color (opacity) denotes increasing
  temperature of the formation epoch: $10^{22}K$ is the end of inflation
  and $10^{11}K$ is big bang nucleosynthesis.  Formation is assumed to
  be instantaneous and monochromatic at the indicated epochs.}
\resizebox{\linewidth}{!}{\includegraphics{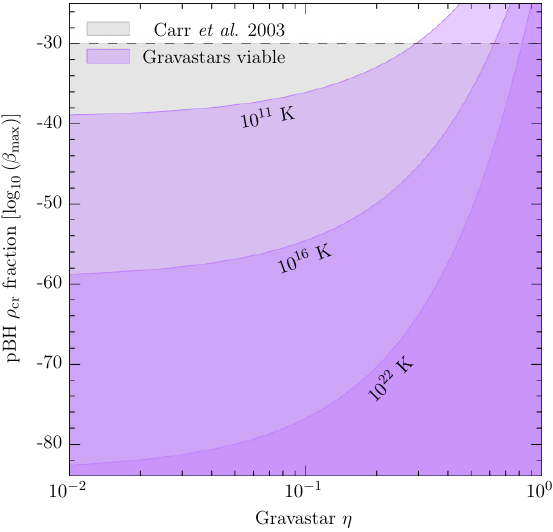}}
\end{figure}

The usual constraint on the fraction $\beta$ of critical density
capable of collapse into a primordial BH is expressed in terms of the
mass of the target hole: $\beta(M)$ \citep[e.g.][]{Carr2003}.  The
mass of the hole is related to $T$ via
\begin{align}
M(a) \simeq \frac{a^2}{2\sqrt{\Omega_r}} = \frac{1}{2\sqrt{\Omega_R}}\left(\frac{T_0}{T}\right)^2
\end{align}
where $\Omega_R$ is the radiation density parameter today.  Thus, we
may construct an analogous constraint $\beta(T)$, where $T$ is some
temperature during the radiation-dominated epoch.  We have switched to
temperature using $a = T_0/T$, where $T_0 = 2.75$K is the CMB
temperature today.

Assume that all primordial BHs are produced at a single moment in time
$a_p$
\begin{align}
\frac{\rmd \Delta_\mathrm{BH}}{\rmd a} = \Omega_\mathrm{pBH}(a) \delta(a - a_p). \label{eqn:pBH_ansatz}
\end{align}
Here $\Omega_\mathrm{pBH}$ is the comoving pBH density at $a_p$.  For
simplicity, consider $\eta$ fixed in time.  In this setting,
Eqn.~(\ref{eqn:kappa}) becomes
\begin{align}
\kappa = a^{-3(1 - \eta)}.
\end{align}
Inserting this and Eqn.~(\ref{eqn:pBH_ansatz}) into Eqn.~(\ref{eqn:prediction}) we find
\begin{align}
\rho_s = \left(\frac{a}{a_p}\right)^{3(1-\eta)}\Omega_\mathrm{pBH}(a_p), 
\end{align}
and evaluating at the present day gives
\begin{align}
\Omega_\Lambda \geqslant a_p^{3(\eta-1)}\Omega_\mathrm{pBH}(a_p). \label{eqn:gstar_pbh_constraint}
\end{align}
The present-day dark energy density gives a conservative upper bound
because we are neglecting accretion.  According to
\citet[][Eqn.~7]{Carr2003},
\begin{align}
\Omega_\mathrm{pBH}(a) = \frac{\beta \Omega_R}{a},
\end{align}
which we can relate to $\Omega_m$ at the epoch of equality $\Omega_R =
a_\mathrm{eq}\Omega_m$.  With these relations,
Eqn.~(\ref{eqn:gstar_pbh_constraint}) becomes
\begin{align}
\Omega_\Lambda \geqslant \frac{\Omega_m a_\mathrm{eq}\beta}{a_p^{4-3\eta}}.
\end{align}
Rearranging the previous expression for $\beta$ we find
\begin{align}
\beta \leqslant \frac{\Omega_\Lambda a_p^{4-3\eta}}{\Omega_m a_\mathrm{eq}} \simeq a_p^{4-3\eta} \times 10^3 \label{eqn:pbh_beta_constraint},
\end{align}
which is presented in Figure~\ref{fig:primordial}.

Of course we do not claim that detection of even a single gravastar
could enforce the constraints shown in Figure~\ref{fig:primordial}.
The essential point, however, is that primordial production of
gravastars is heavily disfavored.  For small $\eta$, the amplification
in energy density from the $a^{3\eta - 4}$ term is extreme at
primordial epochs.

\subsection{Gravastars as the only source of dark energy}
\label{sec:darkages}
The prediction given in Eqn.~(\ref{eqn:prediction}) tightly correlates
the dark energy density to the matter density.  This suggests a
natural resolution to the coincidence problem: gravastars are
responsible for all of the cosmological dark energy.  This possibility
is plausible because the $\kappa(a, a_c) \simeq a^{-3}$ in
Eqn.~(\ref{eqn:prediction}) strongly amplifies the effect of
conversion at early times.  Unfortunately, until gravitational wave
observatories begin to constrain $\Delta_\mathrm{BH}$, a precise
discussion must remain focused on the late-time changes in
$\Omega_\Lambda^\mathrm{eff}$.  To frame these changes in an
appropriate context though, it is useful to understand the general
features of any gravastar formation history that could suffice to
resolve the coincidence problem.  Since this is a very important
consequence of a gravastar population, we will approach the question
in two complementary ways.

\subsubsection{Technique I: Instantaneous formation epoch from present-day BH density}
\label{sec:instantaneous}
\begin{figure}[t]
\caption{\label{fig:instantaneous-formation} Estimated instantaneous
  gravastar formation redshift $z_f$, consistent with present day DE
  density.  The horizontal axis gives the deviation $\eta$ from a
  perfect de Sitter ($w_s = -1$) contribution to the cosmological
  source, due to the presence of gravastar crusts.  Note that the
  onset of star formation $z_\mathrm{early} \equiv 20$ encloses the
  viable region (green/light grey) for $\eta < 0.16$.  The red (medium 
  grey) region indicates $\Omega_\Lambda$ within a factor of 1.5 of
  the present day value, while the blue (dark grey) region indicates
  $\Omega_\Lambda$ within a factor of 0.5 of the present day value.}
\resizebox{\linewidth}{!}{\includegraphics{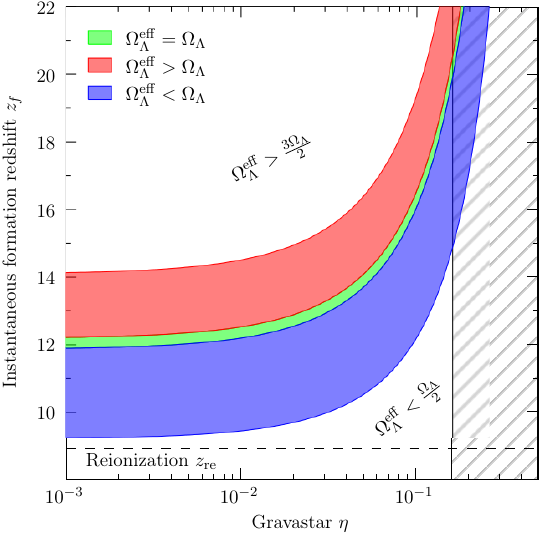}}
\end{figure}

First, we will determine an instantaneous formation epoch, given the
present-day BH density.  In other words, we will determine an
instantaneous formation time $a_f$ for all BH density such that the
correct $\Omega_\Lambda$ is obtained.  When gravitational wave
observatories have constrained this density directly, this estimate
can be checked anew.  Until then, we will use the stellar population
based BH model developed in Appendix~\ref{sec:distribution}.

To proceed with an order of magnitude estimate, we again assume that
all gravastars are produced at a single moment in time $a_f$
\begin{align}
\frac{\Delta_\mathrm{BH}}{\rmd a} = \Omega_\mathrm{BH} \delta(a - a_f). \label{eqn:kaiser_approx}
\end{align}
We take the total cosmological density in gravastars to be the
present-day values
\begin{align}
\Omega_\mathrm{BH} \equiv
\begin{cases}
3.02\times 10^{-4} & \text{(rapid)} \\
3.25\times 10^{-4} & \text{(delayed)}
\end{cases}.
\end{align}
Here, rapid and delayed refer to the stellar collapse model as
detailed by \citet{fryer2012compact}.  These values are likely slight
underestimates due to accretion, which is neglected in
Appendix~\ref{sec:distribution}.  Note that these values are
consistent with the fraction of stars that collapse to BH, as
estimated by \citet{BetheBrown1994}.  Substituting the approximation
Eqn.~(\ref{eqn:kaiser_approx}) into Eqn.~(\ref{eqn:prediction}) we
find
\begin{align}
z_f = \left(\frac{\Omega_\Lambda}{\Omega_\mathrm{BH}}\right)^{1/3(1-\eta)} - 1, \label{eqn:instant_formation_z}
\end{align}
where we have again fixed $\eta$, and used that $\kappa(1,a_c) = 1$
for any fixed $\eta$.  The instantaneous formation epoch is displayed
in Figure~\ref{fig:instantaneous-formation}.

\subsubsection{Technique II: Constant formation from onset, rate consistent with $\gamma$-ray opacity}
\begin{deluxetable}{lcc}
\tablewidth{0pt}  
\tabletypesize{\footnotesize}
\tablecaption{\label{tbl:darkages}Constraints on Population III star formation}
\tablehead{\colhead{Parameter} & \colhead{Value (natural units)} & \colhead{Reference}}
\startdata
$\rmd \rho_*/\rmd t,~z \sim 20$ & $< 5.4\times10^{-2}$ & \citet[][Fig.~2]{inoue2014upper} \\
$\rmd \rho_*/\rmd t,~z \sim 6$ & $< 5.4\times10^{-3}$ & \citet[][Fig.~2]{inoue2014upper} \\
$A$ & $< 22.7$ & \citet{abel2002formation} \\
$\Xi$ & $10^{-1}$ & Table~\ref{tbl:collapse_fractions} \\
\enddata

\tablecomments{data are used to constrain an approximate gravastar
  formation epoch, which suffices to produce all present-day dark
  energy density $\Omega_\Lambda$ by some cutoff redshift $z_f$.
  Variables defined in the text.  The results of this analysis can be
  found in Figure~\ref{fig:cutoff-formation}.}
\end{deluxetable}

\begin{figure}[t]
\caption{\label{fig:cutoff-formation}Estimated dark energy density
  induced by gravastar formation from onset of stellar formation at
  $z_\mathrm{re} \equiv 20$ until some cutoff $z_f$.  This assumes a
  constant, but conservative, primordial rate of stellar density
  production $\rmd \rho/\rmd t \equiv 5.3 \times 10^{-3}$.  This value
  is consistent with the more conservative $\gamma$-ray opacity
  constraints reported by \citet{inoue2014upper} at $z\sim 6$ (see
  Table~\ref{tbl:darkages}).  Further assumptions include a fixed
  fraction of stellar density collapsing into BH and a constant
  amplification due to accretion.  Note consistency with
  Figure~\ref{fig:instantaneous-formation}, and convergence in the de
  Sitter limit $\eta \to 0$.  Green (light grey) indicates production
  of the present-day $\Omega_\Lambda$, red (medium grey) indicates
  excessive production up to $3\Omega_\Lambda/2$, and blue (dark grey)
  indicates insufficient production no less than $\Omega_\Lambda/2$.}
\resizebox{\linewidth}{!}{\includegraphics{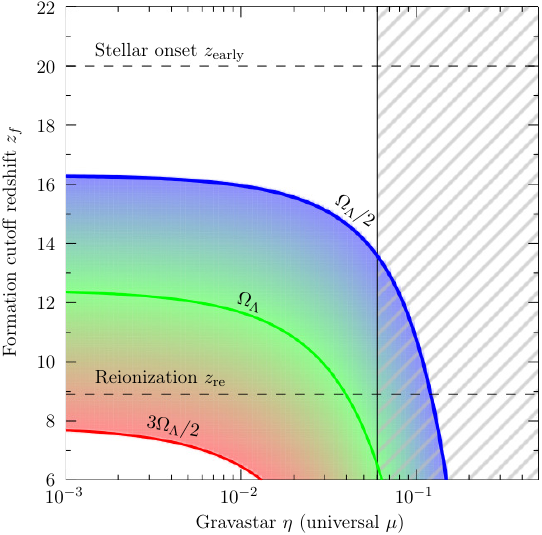}}
\end{figure}
The stellar model used in \S\ref{sec:instantaneous} may suffer from
systematics associated with interpolating the collapse remnant
distributions between only two metallicities.  To estimate the effect
of this systematic, we will use unrelated astrophysical constraints
determined from $\gamma$-ray opacity reported by
\citet{inoue2014upper}.  We will assume that a constant fraction of
stellar density collapses to BH between stellar onset
$z_\mathrm{early}$ up until some cutoff time $z_f$.  We will determine
the value of $z_f$ such that the produced BH density yields the
observed $\Omega_\Lambda$ today.  We will further assume that the
stellar density formation rate during $z_f < z < z_\mathrm{early}$ is
constant.

Since we are considering formation over an extended period of
time, for simplicity, we continue to restrict our attention to $\eta$
fixed in time.  Under these assumptions, we find that
\begin{align}
\Omega_\Lambda = \int_{a_\mathrm{early}}^{a_f} a^{-3(1-\eta)}A\Xi\frac{\rmd \rho_g}{\rmd a}~\rmd a.
\end{align}
Here we have introduced a constant collapse fraction $\Xi$ and a
factor $A$ to account for accretion processes.  We will take $A \sim
10$.  This is reasonable, given that \citet{LiAccretion2006} have
shown that accretion can be very efficient at primordial times. With
the constant star formation rate assumption, we may use the chain rule
to write
\begin{align}
\Omega_\Lambda = A\Xi\frac{\rmd \rho_g}{\rmd t}\int_{a_\mathrm{early}}^{a_f} \frac{\rmd a}{Ha^{4-3\eta}}. \label{eqn:technique2}
\end{align}
We present reasonable data for the above parameters in
Table~\ref{tbl:darkages}, where the formation rates listed are upper
limits.  Given the matter-dominated Hubble factor
\begin{align}
H = \frac{\sqrt{\Omega_m}}{a^{3/2}}, 
\end{align}
we may integrate Eqn.~(\ref{eqn:technique2}) and solve for the cutoff
time.  The result is
\begin{align}
a_f = \left[\frac{3\Omega_\Lambda\sqrt{\Omega_m}}{A\Xi}\left(\frac{\rmd\rho_g}{\rmd t}\right)^{-1}\left(\eta - \frac{1}{2}\right) + a_\mathrm{early}^{3(\eta - 1/2)}\right]^{1/3(\eta - 1/2)} \label{eqn:cutoff_scale}.
\end{align}
The behavior of Eqn.~(\ref{eqn:cutoff_scale}), converted to redshift,
is shown in Figure~\ref{fig:cutoff-formation}.  

Note that we have taken the $10\times$ more conservative bound
established at $z\sim 6$.  If we take $\Xi \sim 0.01$, there is no
viable space.  This is consistent with the stellar model of
Appendix~\ref{sec:distribution}, where the primordial collapse
fraction is $\Xi \sim 0.1$.  It is not surprising that the collapse
remnant model's primordial $\Xi$ is larger than the value of
\citet{BetheBrown1994}, because their analysis depends on iron cores,
which are not present in Population III stars.  Note that the fraction
of matter density converted to gravastar material under our
assumptions,
\begin{align}
\Omega_\mathrm{BH} = A\Xi\frac{\rmd \rho_g}{\rmd t} \int_{a_\mathrm{early}}^{a_f} \frac{\rmd a}{Ha} < 6.5\times10^{-4},
\end{align}
is consistent with the present-day $\Omega_\mathrm{BH}$ used in
\S\ref{sec:instantaneous}.

\subsubsection{Discussion}
\label{sec:nutdiscussion}
In this section, we have considered whether gravastars alone could
account for the present-day observed DE density.  This is plausible
because the DE induced by gravastar formation is amplified by $\sim
1/a^3$.  Using two complementary techniques, we have estimated an
approximate formation epoch for gravastars, which suffices to produce
the present-day observed $\Omega_\Lambda$.  Encouragingly, both
techniques agree that sufficient production is viable over a large
range of $\eta$.  This resolves the coincidence
problem \citep[e.g.][\S6.4]{AmendolaTsujikawa} of a narrow permissible value
for $\Omega_\Lambda$.  Furthermore, both techniques converge toward
$z_f \sim 12$ and produce excluded regions for gravastar $\eta$
\begin{align}
\eta &\lesssim 1.6\times 10^{-1} \qquad \text{(instantaneous)} \\
\eta &\lesssim 6.0\times 10^{-2} \qquad \text{(constant cutoff)}.
\end{align}
Above these thresholds, resolution of the coincidence problem by
gravastars is disfavored.  As we will soon seen, this exclusion is
consistent with those based on late-time behavior and existing Planck
constraint.  

This result is very useful for many reasons.  Most importantly, the
viable region lies squarely in the middle of the epoch of primordial
star formation.  This suggests that a naturally emerging population of
gravastars can produce the correct DE density.  In addition, the
density $\Omega_\mathrm{BH}$ converted to induce this effect is $\sim
0.1\%$ of $\Omega_m$.  This is well within uncertainties on the Planck
best-fit value for $\Omega_m$.  Since $z_f \ll z_\mathrm{CMB}$,
production of a sufficient population can occur without breaking
agreement with precision CMB astronomy.  In other words, CMB
anisotropies are not altered at any level.  Finally, gravastar
formation at $z < 20$ can establish a present-day valued DE density
after star formation has begun.  Thus, initial conditions, and
therefore results, of precision $N$-body simulations are not altered.

In any scenario where $\eta > 0$, the produced physical DE density
will diminish with time $\propto 1/a^{3\eta}$.  Relative to the
required present day value of $\Omega_\Lambda$ then, the physical DE
density is larger in the past.  This behavior is not relevant for the
present discussion, as it represents a minor increase in any physical
DE density due to gravastars with viable $\eta$.  During the dark
ages, the matter density (dominated by dark matter) completely overwhelms any
DE.  This effect, however, is tracked in
\S\ref{sec:planck_constraint}.

For the subsequent discussion, we interpret these estimates in the
following way: \emph{if} gravastars are to resolve the coincidence
problem, we expect that most of the dark energy density will be
established by stellar and accretion physics taking place during $5 <
z < 20$.

\subsection{Gravastars analyzed in the dark fluid framework}
\label{sec:planck_constraint}
\begin{figure}[t]
\caption{\label{fig:weff_universal_mu}Time-evolution of gravastar
  $w_\mathrm{eff}(a)$ estimated from the stellar population.  Both
  rapid (orange/light grey) and delayed (blue/dark grey) stellar
  collapse scenarios are shown.  Best-fit lines (dashed) assuming the
  Planck linear ansatz near $a=1$, shown for $\eta = 3\times 10^{-2}$
  only.  Colors (greyshades) indicate distinct $\eta$
  (c.f. Figure~\ref{fig:planck-consistency-eta}).  Note breakdown of
  the dark fluid (linear) approximation at $a=0.9$.}
\resizebox{\linewidth}{!}{\includegraphics{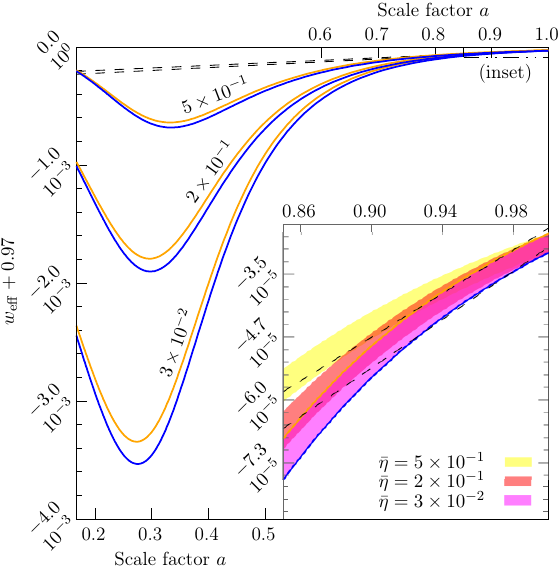}}
\end{figure}

Dark Energy is most commonly studied through constraint of three
parameters: an energy density $\Omega_\Lambda$, and the linear Taylor
expansion, about $a\equiv 1$, of its equation of state parameter
$w(a)$.  This analysis of possible deviations from $\Lambda$CDM comes
from the anticipated dynamics of a minimally coupled scalar field: the
dynamics of an internally conserved fluid.

In the gravastar scenario, stellar collapse rates, subsequent
accretion, and evolution in the stellar remnant distribution all
induce changes in the DE density $\rho_s$.  We can, however, choose to
interpret this time-dependence of $\rho_s$ as the dynamics of an
internally conserved fluid.  This leads to an effective equation of
state parameter as follows.  Define $w_\mathrm{eff}$ through
\begin{align}
\derp{\rho_s} \equiv -3w_\mathrm{eff}H\rho_s. \label{eqn:weff_def}
\end{align}
Subtracting Eqn.~(\ref{eqn:cons_bh}) from Eqn.~(\ref{eqn:weff_def}),
switching to scale factor, substituting Eqn.~(\ref{eqn:prediction}),
and solving for $w_\mathrm{eff}$ gives
\begin{align}
w_\mathrm{eff} = w_s - \frac{a\Delta_\mathrm{BH}'}{3\rho_s}
\end{align}
where prime denotes derivative with respect to $a$.  Using the
definition of $\kappa$ in Eqn.~(\ref{eqn:kappa}) and linearity of the
integral, we may re-express this relative to the present-day value of
$\Omega_\Lambda$
\begin{align}
w_\mathrm{eff} = w_s - \frac{\rmd \Delta_\mathrm{BH}}{\rmd a}\kappa(a,1)\frac{a}{3}\left[\Omega_\Lambda - \int_a^1 \frac{\rmd \Delta_\mathrm{BH}}{\rmd a'}\kappa(a',1)~\rmd a'\right]^{-1}. \label{eqn:weff_lowz}
\end{align}
This form is advantageous as it samples only values very near to the
present epoch, where constraint and systematics are
well-characterized.  The first-order Taylor expansion in $(1-a)$ of
Eqn.~(\ref{eqn:weff_lowz}) does not contain $\kappa$ at all
\begin{align}
\xo{(0)}{w_\mathrm{eff}} &= -1 + \left(\eta - \frac{\Delta_\mathrm{BH}'}{3\Omega_\Lambda}\right)\Bigg|_{a=1} \label{eqn:weff_0} \\
\xo{(1)}{w_\mathrm{eff}} &= \frac{ \Omega_\Lambda(\Delta_\mathrm{BH}'' - 2\Delta_\mathrm{BH}') - \Delta_\mathrm{BH}'^2}{3\Omega_\Lambda^2}\Bigg|_{a=1} + \left(\frac{\eta \Delta_\mathrm{BH}'}{\Omega_\Lambda} - \eta'\right)\Bigg|_{a=1}. \label{eqn:weff_1}
\end{align}
The coefficient Eqns.~(\ref{eqn:weff_0}) and (\ref{eqn:weff_1}),
combined with astrophysically measured values and uncertainties in BH
population parameters, will give a region which can be immediately
compared against any Planck-style $w_0-w_a$ constraint diagram.

The results of \S\ref{sec:darkages} suggest that most of
$\Omega_\Lambda$ should be already produced by $z \sim 5$.  In this
case, we may approximate Eqn.~(\ref{eqn:weff_lowz}) as
\begin{align}
w_\mathrm{eff} \simeq w_s - \Delta_\mathrm{BH}'\kappa(a,1)\frac{a}{3\Omega_\Lambda}\left[1 - \frac{\Delta_\mathrm{BH}'\kappa(a,1)}{\Omega_\Lambda}\right] \label{eqn:weff_lowz_expansion}.
\end{align}

We have shown that knowledge of the black hole mass function and the
present-day dark energy density completely determines
$w_\mathrm{eff}$.  In this way, gravitational wave observatory
measurements of the BH population make definitive cosmological
predictions.  We already have a wealth of cosmological observations,
however, with upcoming experiments \citep[e.g.][DESI]{levi2013desi}
set to constrain $w_\mathrm{eff}$.  We may thus make predictions for
gravitational wave observatories by inverting the above procedure.
Solving for $\rmd \Delta_\mathrm{BH}/\rmd a$ by differentiating
Eqn.~(\ref{eqn:weff_lowz}) and re-integrating, we find
\begin{align}
\frac{\rmd \Delta_\mathrm{BH}}{\rmd a} &= 3\Omega_\Lambda\frac{w_s(a) - w_\mathrm{eff}(a)}{a}\exp\left[3\int_a^1 w_\mathrm{eff}(a') \frac{\rmd a'}{a'}\right]\label{eqn:lowz_BHprediction_mufree}. 
\end{align}
Note that we have removed $\kappa$ entirely through use of
Eqn.~(\ref{eqn:kappa}).
Given the particularly simple $w_\mathrm{eff}$ assumed by Planck
\begin{align}
w_\mathrm{eff}(a) \equiv w_0 + w_a(1-a), \label{eqn:planck_w}
\end{align} 
we may express Eqn.~(\ref{eqn:lowz_BHprediction_mufree}) in closed form
\begin{align}
\frac{\rmd \Delta_\mathrm{BH}}{\rmd a} &= 3\Omega_\Lambda\left[\eta - (1 + w_0 + w_a) + w_aa\right]\frac{\exp\left[3w_a(a-1)\right]}{a^{3(w_0 + w_a) + 1}}.\label{eqn:dBHda}
\end{align}

\subsubsection{Estimation of $w_\mathrm{eff}$ from the comoving stellar density}
In the gravastar scenario, the physical origin of DE is completely
different than the internal dynamics of a scalar field.  The linear
model Eqn.~(\ref{eqn:planck_w}), based on the dark fluid assumption,
may not be the most suitable for characterization of a late-time
gravastar DE contribution.  In this section, we use a BH population
model, developed from the stellar population in
Appendix~\ref{sec:distribution}, to investigate $w_\mathrm{eff}$ in
the gravastar scenario.  These results will allow us to estimate when
Eqns.~(\ref{eqn:planck_w}) and (\ref{eqn:dBHda}) well-approximate the
time-evolution of the comoving BH density.

The BH population model of Appendix~\ref{sec:distribution} does not
account for accretion, so we must remain outside of any epoch where BH
accretion is significant.  Consistent with the simulations of
\citet{LiAccretion2006}, we consider only $z < 5$.  To begin
evaluation of Eqn.~(\ref{eqn:weff_lowz}), we require $\kappa(a,1)$,
which requires $\eta$.  For fixed $\eta$, the required integrations
become trivial.  We find
\begin{align}
\kappa(a,1) &= \frac{1}{a^{3(1-\eta)}}
\end{align}
and predict for Eqn.~(\ref{eqn:weff_lowz_expansion})
\begin{align}
w_\mathrm{eff} \simeq -1 + \eta - \frac{\Delta_\mathrm{BH}'}{3\Omega_\Lambda a^{2 - 3\eta}}\left[1 - \frac{\Delta_\mathrm{BH}'}{\Omega_\Lambda a^{3(\eta - 1)}}\right] \label{eqn:weff_approximation}.
\end{align}
This relation is displayed in Figure~\ref{fig:weff_universal_mu}.  The
range corresponds to ``rapid'' and ``delayed'' models of stellar
collapse, as determined by \citet{fryer2012compact}.  The collapse
model determines the distribution of remnant masses, which evolves in
time with metallicity.  Given $\Delta_\mathrm{BH}$,
Eqn.~(\ref{eqn:weff_approximation}) provides a single parameter fit,
with sub-percent precision, viable for $z < 5$.  It is clear that the
Planck linear ansatz is only useful for
\begin{align}
a > 0.9 \qquad z < 0.11
\end{align}
with departures growing significantly worse as $\eta \to 0$.

\subsubsection{Planck and DES constraint of $\eta$ at the present epoch}
\begin{figure}[t]
\caption{\label{fig:planck-consistency-eta}Gravastar constraint region
  (orange line/horizontal axis) estimated from the stellar population.
  Region is superimposed upon Planck constraints of the Dark Energy
  equation of state evolution.  Overlaid ``chevrons'' represent
  positive BH density constraints for $a > 0.9$, which is the
  lower-bound $a$ for validity of the Taylor expansion.  Colors
  (greyshades) indicate distinct $\eta$
  (c.f. Figure~\ref{fig:weff_universal_mu}).  Chevron boundaries have
  thickness $\Delta_\mathrm{BH}'(1)/3\Omega_\Lambda \sim 10^{-5}$.
  The intersection of these vertical bands with the horizontal band
  gives viable $w_0 - w_a$ for any particular $\eta$
  (c.f. Figure~{\ref{fig:planck-consistency-zoomed}}).  Note that
  Planck+WBR is the Planck+WL+BAO/RSD combined constraint, where
  WL is weak lensing and BAO/RSD is baryon acoustic oscillations/redshift
  space distortions.  BSH is a combined constraint from BAO, Type Ia
  Supernovae, and local measurements of the Hubble constant.}
\resizebox{\linewidth}{!}{\includegraphics{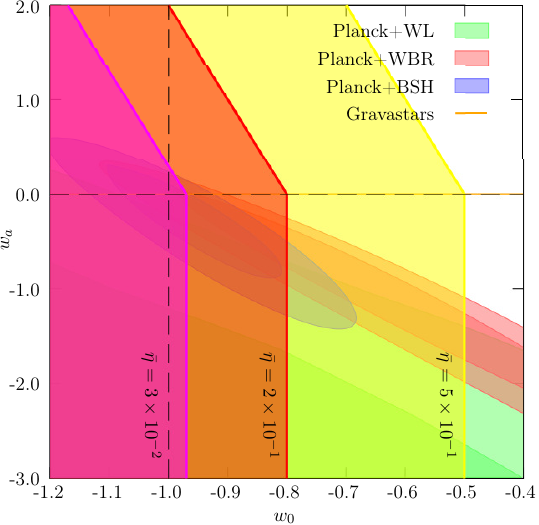}}
\end{figure}

\begin{figure}[t]
\caption{\label{fig:planck-consistency-zoomed} Gravastar constraint
  region (orange/light grey) estimated from the stellar population, vertical axis
  magnified by $10^5$ compared to
  Figure~\ref{fig:planck-consistency-eta}.  The region is superimposed
  upon Planck constraints of the Dark Energy equation of state
  evolution.  Stellar collapse model indicated via label.  Acronyms
  are defined in the caption of
  Figure~\ref{fig:planck-consistency-eta}.  Planck $2\sigma$ contours
  have been removed for clarity.  For illustration, permissible region
  for $\eta \equiv 3\times 10^{-2}$ shown, with thickness $10^{5}$
  smaller than indicated.  The permissible region is the overlap of
  the orange band and the vertical orange line.}
\resizebox{\linewidth}{!}{\includegraphics{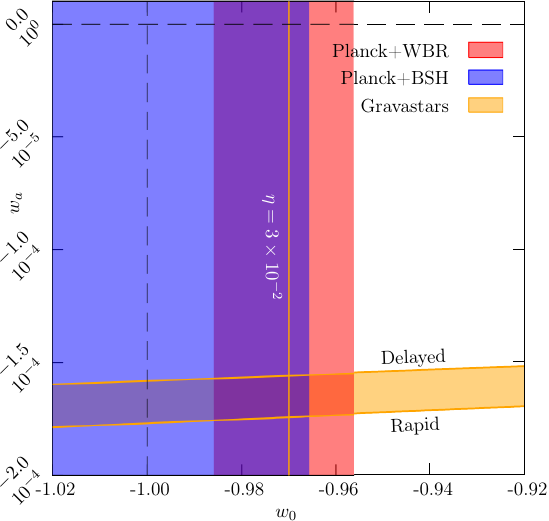}}
\end{figure}

Physically, $\rmd \Delta_\mathrm{BH}/\rmd a \geqslant 0$ must be
satisfied wherever ansatz Eqn.~(\ref{eqn:planck_w}) is valid.  From
Eqn.~(\ref{eqn:lowz_BHprediction_mufree}), this constraint becomes an
upper bound
\begin{align}
\xo{(0)}{w_\mathrm{eff}} \leqslant \eta(1) - 1. \label{eqn:positivity_bounds}
\end{align}
Regardless of any time-evolution in $\eta$, we may use
Eqns.~(\ref{eqn:weff_0}) and (\ref{eqn:weff_1}) to constrain $\eta$
now.  Eliminating $\eta$ from these equations, we find a permissible
line through $w_0-w_a$ space
\begin{align}
\xo{(1)}{w_\mathrm{eff}} = \frac{\Delta_\mathrm{BH}'}{\Omega_\Lambda}\xo{(0)}{w_\mathrm{eff}} + \left(\frac{\Delta_\mathrm{BH}'' + \Delta_\mathrm{BH}'}{3\Omega_\Lambda} - \eta'(1)\right) \label{eqn:planck_constraint_line}.
\end{align}
We will use the BH population model, developed from the stellar
population in Appendix~\ref{sec:distribution}, to estimate the
derivatives of $\Delta_\mathrm{BH}$.  From
Eqn.~(\ref{eqn:stellar_bh_model}), we numerically find that
\begin{align}
\Delta_\mathrm{BH}'(1) &= \begin{cases}
5.364 \times 10^{-5} & \text{(rapid}) \\
6.135 \times 10^{-5} & \text{(delayed})
\end{cases} \\
\Delta_\mathrm{BH}''(1) &= \begin{cases}
-2.214 \times 10^{-4} & \text{(rapid)} \\
-2.447 \times 10^{-4} & \text{(delayed)} 
\end{cases},
\end{align}
which will produce a permissible band in $w_0-w_a$ space.  For
$\eta' \equiv 0$, this band is displayed in Figures
\ref{fig:planck-consistency-eta} and
\ref{fig:planck-consistency-zoomed}.  Evidently, Planck data
disfavor gravastars with large $\eta$
\begin{align}
0 < \eta(1) \leqslant 3\times10^{-2} \qquad \eta'(1) \equiv 0.
\end{align}
Within this range, however, gravastars are consistent with
Planck best fit constraints.  

By inspection of Eqn.~(\ref{eqn:planck_constraint_line}),
$\eta'$ simply translates the constraint region vertically.  If
we are to remain consistent with $\rmd \Delta_\mathrm{BH}/\rmd a
\geqslant 0$ for $0.9 < a < 1$, we must have
\begin{align}
\xo{(1)}{w_\mathrm{eff}} \leqslant -10\xo{(0)}{w_\mathrm{eff}} + 10[\eta(a) - 1].
\end{align}
This bound, and the positivity bounds given in
Eqn.~(\ref{eqn:positivity_bounds}) are displayed in
Figure~\ref{fig:planck-consistency-eta} for a variety of $\eta$.

Recent results constraining the $w$CDM model from the Dark Energy
Survey (DES) ~\citep{abbott2017dark} cannot be immediately applied to
constrain $\eta$.  This is because $w_\mathrm{eff}$ induced by a
gravastar population changes in time even if $\eta$ remains fixed for
all time.  Since we predict only small changes in $w_\mathrm{eff}$,
however, it is reasonable to expect that the $w$CDM model will
approximate the gravastar scenario at late times.  Indeed, their
reported value of $w$ \citep[][Eqn.~VII.5]{abbott2017dark} becomes
the following constraint on $\eta$
\begin{align}
\eta < 4\times10^{-2} \qquad (w_0 =-1.00^{+0.04}_{-0.05}), 
\end{align}
which is consistent with the constraints reported in
\S\ref{sec:nutdiscussion}.

\section{Conclusion}
Gravitational vacuum stars have become a viable and popular
theoretical alternative to the pathological classical black hole (BH)
solutions of GR.  These objects appear as BHs to exterior vacuum
observers, but contain de Sitter interiors beneath a thin crust.  This
crust is located above the classical Schwarzchild horizon, placing the
entire gravastar in causal contact with the exterior universe.  Many
existing studies have focused on the observational consequences of the
crust region for optical and gravitational wave signatures.  These
studies are hindered by systematics relating to the crust, which is
only loosely constrained theoretically.  In this paper, we have
developed complementary constraints on the gravastar scenario based on
the well-established properties of their de Sitter interiors.

We place our gravastars within a flat Friedmann cosmology, which is
nowhere vacuum, and show from the action principle that the zero-order
Friedmann source must contain an averaged term sensitive to the
interiors.  This is consistent with Birkhoff's theorem, which does not
apply without vacuum boundaries.  Through conservation of
stress-energy, this term induces a time-dependent Dark Energy (DE)
density.  This density is directly correlated to the evolution of
non-linear structure via star formation and subsequent collapse.  The
gravastar crust produces a small deviation $\eta$ from a pure de
Sitter ($w=-1$) equation of state.  This deviation becomes the single
parameter characterizing the gravastar population.

We replace all black holes with gravastars and consider the
cosmological effects of their subsequent DE contribution at three
epochs.  During the primordial epoch ($T\sim10^{22}\mathrm{K}-T\sim
10^{11}\mathrm{K}$), we find that the fraction of matter collapsing
into a primordial gravastar population with $\eta < 10^{-1}$ is
constrained between $\sim 10-50$ orders of magnitude more than
existing primordial BH constraints.  During the dark ages $(8.9
\lesssim z \lesssim 20)$ we show, via two approaches, that existing
astrophysical data support formation of a population of gravastars
that can account for all of the present-day DE density.  We show that
any gravastar population with crust parameter $\eta < 6\times 10^{-2}$
can resolve the coincidence problem.  During late-times ($z < 5$), we
precisely interpret the gravastar scenario in the usual language of a
time-varying dark fluid.  Using a BH population model built from the
cosmic star formation history and stellar collapse simulations, we
predict time-variation in the magnitude of $w(a)$ that tracks star
formation.  We demonstrate complete consistency with Planck, given a
gravastar population with $\eta < 3\times 10^{-2}$.  Further, we
predict very little time-variation in $w(a)$ at late times, consistent
with the recent results of the Dark Energy Survey.

We make definitive predictions for both the gravitational astronomy
community and dark energy surveys in the form of unexpected
quantitative correlations between the time-evolution of the DE density
and the BH population.  In summary, the cosmological consequences of a
gravastar population are unambiguous, readily testable, and already
resolve many outstanding observational questions, without requiring
any \emph{ad hoc} departure from GR.

All code for generating the presented data and its
visualizations is released publicly (oprcp4).

\software{
  \href{https://github.com/kcroker/oprcp4}{oprcp4},
  \href{https://www.scipy.org}{\texttt{scipy}}~\citep{scipy}, 
  \href{http://maxima.sourceforge.net/}{GNU Maxima},
  \href{http://www.gnuplot.info}{gnuplot}}

\acknowledgments 

This paper is dedicated to the memory of Prof. J. M. J. Madey,
inventor of the free-electron laser.  His emphasis on the ``paramount
importance of boundary conditions'' heavily influenced this research.
The author thanks N. Kaiser (IfA) for sustained theoretical criticism,
J. Weiner (U.~Hawai`i) for thorough technical feedback concerning the
action, and T. Browder (U.~Hawai`i) and K. Nishimura (U.~Hawai`i) for
copious feedback during the preparation of all versions of the
manuscript.  Additional thanks go to S. Ballmer (aLIGO) for
conversations concerning the capabilities of present and planned
gravitational wave observatories, C. Corti (AMS02) for visualization
suggestions, C. McPartland (IfA) for guidance in the stellar
literature, N. Warrington (U. Maryland) for stimulating discussions
and feedback, J. Kuhn (IfA) for comments on rich clusters, J. Learned
(U.~Hawai`i) for encouragement, R. Matsuda (U.~Tokyo/IPMU) for
comments on clarity, and The University of Tokyo for hospitality
during the preparation of this manuscript.  This work was performed
with financial support from the Fulbright U.S. Student Program.

\appendix
\section{Assumptions of the \emph{de facto} Friedmann source at late-times}
\label{sec:assumptions}
In this appendix, we highlight non-trivial assumptions that enter the
\emph{de facto} Friedmann source (DFFS) during matter domination.  In
the DFFS, one imposes an additional hypothesis on the physical
background matter density at late-times~\citep[][\S9,
  \S97]{peebles1980large}
\begin{align}
\bar{\rho}_b(t) \propto a^{-3} \label{eqn:ass_constant}.
\end{align}
To understand its origin, consider the conservation of stress-energy
statement for a single component perfect fluid
\begin{align}
\frac{\rmd}{\rmd t} \left(\bar{\rho}_b a^3\right) = -3H\bar{\mathcal{P}}_ba^3.
\end{align}
Here $H$ is the Hubble parameter (defined in \S\ref{sec:friedmann})
and $\bar{\mathcal{P}}_b \equiv T^{kk}$ is the background physical
pressure.  Equation~(\ref{eqn:ass_constant}) then follows if one
\emph{defines} $\bar{\mathcal{P}}_b(t) \equiv 0$.  Note that this is
an assumption beyond Einstein's equations with the RW metric.  The
origin of this assumption is microphysical.  For $20 \lesssim z
\lesssim z_\mathrm{CMB}$, mean free paths of all particles (even
hypothetical dark matter) are very large; particles are not scattering
into or out of comoving volumes.  This fixes the comoving number
density of particles.  Since the rest masses of elementary particles
do not change, it follows that $\bar{\mathcal{P}}_b$ must be zero.

Given the lack of a formal mathematical definition for the DFFS, we
consider the operational definition of the perturbations to
$\bar{\rho}_b(t)$, given by \citet[\S81]{peebles1980large}.  He states
that ``we can imagine that observers spread through the universe and
moving with the matter keep a record of the local density as a
function of proper time, $\bar{\rho}(\mathbf{x}, t)$.  As the observers
come within the horizon, their records can be acquired and compared.''
We now paraphrase this operational procedure:
\begin{enumerate}
\item{Each observer measures and broadcasts (e.g. via light signals with presumably the same frequency)
  their own local density $\bar{\rho}(\mathbf{x}, t)$.}
\item{Each observer receives the others' broadcasts and averages them with his own to produce a background $\bar{\rho}_b(t)$.}
\item{Each observer computes his density perturbation as $\bar{\rho} - \bar{\rho}_b$.}
\end{enumerate}
The same procedure may be performed with the local pressure
$\bar{\mathcal{P}}(\mathbf{x}, t)$, which will be non-zero in general.
Between $20 \lesssim z < z_\mathrm{CMB}$, averaged pressures remain
negligible compared to averaged energy densities.  Below $z \lesssim
20$, however, collapsed structures begin to form. 

In order to justify continued use of the homogeneous and isotropic
fluid ansatz for $z \lesssim 20$, one redefines Peebles' observers.
One regards an observer as reporting on a very large volume containing
many gravitationally bound systems.  Next, one invokes any number of
formal embedding solutions to GR~\citep[e.g.][]{ES1945}, which glue a
Schwarzchild metric into a RW metric.  The existence of such formal
solutions motivates defining a spatially isolated system to contribute
only an active gravitational mass to the averaged energy density
$T^{00}$.  In this setting, the only pressures that could contribute
to $T^{kk}$ would be from large structures, like rich clusters.  As we
show in Appendix~\ref{sec:dispersion}, however, upper bound pressures
for these systems are $\mathcal{P} \lesssim 10^{-5}\rho$.  Thus, it
may seem completely justifiable to continue to regard
$\bar{\mathcal{P}}_b \equiv 0$.

It must be emphasized, however, that formal embedding solutions are
logically distinct from the perturbative treatment.  The perturbative
treatment assumes instead the exact RW background.  Suppose we were
unable to redefine our observers, and consider a Peebles' type
observer within the core of a gravastar.  Here the pressure is equal
in magnitude, but opposite in sign, to the energy density.  The DFFS
discards these contributions.  In other words, the DFFS defines
$\bar{\mathcal{P}}_b \equiv 0$ for $z < 20$ because this is consistent
with Newtonian intuition and static strong gravity in asymptotically
flat space.

\section{The spatially averaged nature of the Friedmann source}
\label{sec:action}
In this Appendix, we rederive Friedmann's equations directly from the
Einstein-Hilbert action.  The action is a manifestly global quantity,
from which local equations of motion are constructed.  Since the
action is an integral statement over the spacetime manifold, it will
allow us to cleanly reconcile the isotropy and homogeneity of the RW
metric ansatz with the anisotropic distribution of actual
stress-energy.  

Our approach will be to rewrite Friedmann's isotropic and homogeneous
model as a scalar theory in flat-space.  The result will be an
unambiguous interpretation of the Friedmann source as the flat-space,
comoving coordinate average of the trace of the stress-tensor.  This
well-motivated notion of ``locality'' within the Friedmann model is
lost if one begins with Einstein's equations under the RW ansatz,
simply because one must invoke some \emph{ad hoc} procedure
(e.g. Appendix~\ref{sec:assumptions}) to produce an isotropic and
homogeneous source from the actual stress-energy.

The principle of stationary action requires that the total action $S$
be stable, at first order, to variations in the dynamical degrees of
freedom
\begin{align}
\delta S \equiv \delta(S_M + S_G) \equiv 0 \label{eqn:deltaaction}
\end{align}
where $S_M$ represents the matter action and $S_G$ represents the
gravitational action.  We will consider the canonical definition of the
stress-energy tensor $\bar{T}_{\mu\nu}$ given by \citet[\S12.2]{WeinbergGR}
\begin{align}
\delta S_M \equiv -\frac{1}{2}\int_\mathcal{M} \bar{T}_{\mu\nu}\delta \bar{g}^{\mu\nu}\sqrt{-\bar{g}}~\rmd^4x. \label{eqn:deltaSM}
\end{align}
Here $\mathcal{M}$ denotes the smooth manifold associated with the
metric $\bar{g}_{\mu\nu}$~\citep[\S3.2]{DaddyONeill}, $\delta
\bar{g}^{\mu\nu}$ is the variation of the inverse metric, and
$\sqrt{-\bar{g}}$ is the metric determinant.  Recall that an overset
bar denotes a physical quantity, as opposed to a comoving one.  For
the gravitational action $S_G$, consider the usual Einstein-Hilbert
action
\begin{align}
S_G \equiv \frac{1}{16\pi G} \int_\mathcal{M} \bar{R}\sqrt{-\bar{g}}~\rmd^4x. \label{eqn:SG}
\end{align}
Here $\bar{R}$ is the Ricci scalar determined from the Levi-Civita
connection compatible with $\bar{g}_{\mu\nu}$ and we are using MKS
units with $c\equiv 1$.  

In comoving coordinates, the RW metric takes the form of
Eqn.~(\ref{eqn:rw}).  Defining the conformal time $\eta$ through
\begin{align}
\rmd t \equiv a(\eta)~\rmd\eta, \label{eqn:conft_defn}
\end{align}
Eqn.~(\ref{eqn:rw}) becomes
\begin{align}
\rmd s^2 = a^2\left(-\rmd \eta^2 + \rmd\mathbf{x}^2\right).
\label{eqn:rw_cflat}
\end{align}
Note that Eqn.~(\ref{eqn:rw_cflat}) takes the form of a conformal
transformation of flat spacetime
\begin{align}
\bar{g}_{\mu\nu} = a(\eta)^2\eta_{\mu\nu} \label{eqn:conf_flat_metric}
\end{align}
where $\eta_{\mu\nu}$ is the Minkowski metric of special relativity.
We may thus express $\bar{R}$ in the conformally flat frame
\citep[\S D]{WaldGR}
\begin{align}
\bar{R} &= a^{-2}R - 2(n-1)\eta^{\mu\nu}a^{-3}\nabla_\mu\nabla_\nu a \qquad (n \equiv 4)\\
&= -6\eta^{\mu\nu}a^{-3}\partial_\mu \partial_\nu a.
\end{align}
with regular partial derivatives.  Similarly, the metric determinant
in the conformally flat frame becomes
\begin{align}
\sqrt{-\bar{g}} = a^4. \label{eqn:jacobian} 
\end{align}
The gravitational action then becomes
\begin{align}
S_G = -\frac{6}{16\pi G} \int_\mathcal{M} a \eta^{\mu\nu} \partial_\mu\partial_\nu a~\rmd^4x
\end{align}
and its variation with respect to the single physical degree of
freedom $a(\eta)$ is
\begin{align}
\delta S_G = -\frac{3}{4\pi G} \int_\mathcal{M} \delta a \partial^\mu \partial_\mu a~\rmd^4x  \label{eqn:deltaSG_conf}
\end{align}
where we have discarded the boundary term. 

To find $\delta S_M$, we start with the most general fluid
stress-tensor, which applies to any type of matter and coordinate
choice.  It is given by \citet[][Eqn.~6]{hu2004covariant} as
\begin{align}
\bar{T}^\rho_\nu \equiv
\begin{bmatrix}
  -\bar{\rho} & \bar{v}_i \\ 
  \bar{u}^j &
  \bar{\pi}^j_i
\end{bmatrix}(\eta, \mathbf{x}). \label{eqn:generic_fluid_stress}
\end{align}
where $\bar{v}^i$ and $\bar{u}^i$ are future-directed timelike vector fields.  We have eliminated the purely
formal separation of background from perturbation for clarity.  Now,
we may lower an index in the usual way
\begin{align}
\bar{T}_{\mu\nu} = \bar{T}^\rho_\nu\bar{g}_{\rho\mu}. \label{eqn:lowering}
\end{align}
and use the chain rule to write 
\begin{align}
\delta \bar{g}^{\mu\nu} = -2a^{-3} \eta^{\mu\nu} \delta a.
\end{align}
Substitution of Eqns.~(\ref{eqn:conf_flat_metric}),
(\ref{eqn:jacobian}), (\ref{eqn:generic_fluid_stress}), and
(\ref{eqn:lowering}) into Eqn.~(\ref{eqn:deltaSM}) then gives
\begin{align}
\delta S_M = \int_\mathcal{M} a^3\left(-\bar{\rho} + \bar{\pi}_1^1 + \bar{\pi}_2^2 + \bar{\pi}_3^3\right)\delta a~\rmd^4x. \label{eqn:deltaSM_conf_expanded}
\end{align}
We may compress this further by defining the following notation
\begin{align}
\bar{\mathcal{P}}(\eta, \mathbf{x}) \equiv \frac{1}{3}\sum_{i=1}^3 \bar{\pi}_i^i(\eta, \mathbf{x})
\end{align}
to arrive at the desired expression
\begin{align}
\delta S_M = \int_\mathcal{M} a^3\left(-\bar{\rho} + 3\bar{\mathcal{P}}\right)\delta a~\rmd^4x.
\label{eqn:deltaSM_conf}
\end{align}

We may now make explicit our claim that the Friedmann source is a
spatial-slice average.  Under the constraints of isotropy and
homogeneity, $\delta a$ can only be a function of $\eta$.  This is
because the variation must be consistent with the constraints imposed
on the system under study~\citep[e.g.][]{LanczosVariational}.  We may
thus expand the iterated integral of Eqn.~(\ref{eqn:deltaSM_conf})
\begin{align}
\delta S_M &= \int a^3 \delta a \int_\mathcal{V} [-\bar{\rho}(\eta, \mathbf{x}) + 3\bar{\mathcal{P}}(\eta, \mathbf{x})]~\rmd^3\mathbf{x}~\rmd \eta \\
&=\int a^3 \delta a \mathcal{V}\left<-\bar{\rho} + 3\bar{\mathcal{P}}\right>~\rmd \eta \label{eqn:deltaSM_final}
\end{align}
where $\mathcal{V}$ is some fiducial spatial volume.  Consider a
typical arbitrary variation without symmetry requirements,
i.e. $\delta g_{\mu\nu}(\eta, \mathbf{x})$.  Since the variation is
arbitrary, one is free to consider variations with compact
spatiotemporal support.  In such a case, the integrals appearing in the
action and its variation can always be regarded as finite.  In the RW
setting, however, we are not free to enforce compact spatial support
for $\delta a(\eta)$.  This means that $\mathcal{V}$ is infinite, and
convergence issues might arise from an iterated integral
representation (i.e. Fubini's theorem).  This can be remedied by
regarding $\mathcal{V}$ as resulting from an arbitrary spatial cutoff
for the integration domain.  One then verifies that the result is
independent of the cutoff.

Indeed, substitution of Eqns.~(\ref{eqn:deltaSG_conf}) and
(\ref{eqn:deltaSM_final}) into Eqn.~(\ref{eqn:deltaaction}) gives
\begin{align}
\int \delta a \left\{ \frac{\rmd^2 a}{\rmd \eta^2} - \frac{4\pi G}{3}a^3\left<\bar{\rho} - 3\bar{\mathcal{P}}\right>\right\}~\rmd
\eta = 0
\end{align}
where the fiducial spatial volume successfully divides off.  The
resulting field equation
\begin{align}
\frac{\rmd^2 a}{\rmd \eta^2} = \frac{4\pi G}{3}a^3\left<\bar{\rho} - 3\bar{\mathcal{P}}\right> \label{eqn:conformal_eom}
\end{align}
agrees with the standard Eqns.~(\ref{eqn:fried1}) and
(\ref{eqn:fried2}) summed and shifted to conformal time.  Note that,
in the standard scenario, all $\bar{\pi}_i^i$ are equal, and so
$\bar{\mathcal{P}}$ is the usual isotropic pressure.  The advantage of working
directly from the action is that the spatial average of the source is
made manifest.  This is the main result of this appendix.

\subsection{Discussion}
The spatial average of Eqn.~(\ref{eqn:conformal_eom}) is entirely
consistent with the operational definition given by
\citet[\S81]{peebles1980large}.  The temporary appearance of the
spatial volume $\mathcal{V}$ is interesting.  Einstein's equations are
``local'' in the sense of ``localized in space.''  This leads to the
intuition that all dynamical fields are position-dependent.  In the
context of cosmology, this is reinforced from the Newtonian
approximation, where $a(\eta, \mathbf{x})$ defines a coordinate change
\citep[e.g.][]{peebles1980large}, which can vary from point to point.
Together, these predispositions have sometimes
\citep[e.g.][]{racz2017concordance} led to the perception that there
is some sort of local scale factor $a(\eta, \mathbf{x})$, which the
Friedmann model somehow digests into $a(\eta)$.

One can decompose Eqn.~(\ref{eqn:deltaSM_conf}) into a sum of $N$
open, disjoint, spacelike regions and average over these
\begin{align}
\delta S_M = \sum_{m\subset\mathcal{M}}^N\int a^3\delta a\mathcal{V}_m\left<-\bar{\rho} + 3\bar{\mathcal{P}}\right>_m~\rmd \eta.
\end{align}
Here, the cutoff procedure discussed previously is equivalent to
requiring $N < \infty$.  Until $\mathcal{V}_m$ is large enough for a
given epoch, these averages will not be equal and so the equations of
motion will not be well-defined.  This clarifies $\mathcal{V}^{1/3}$
as the lower bound length-scale implicit to the isotropic perfect
fluid model~\citep[\S1]{Landau6}.  The novelty within cosmology is
that this length-scale is, in general and practice, time-dependent.
Technically, one should not assume the same perfect fluid model
throughout the entire history of the universe.

\section{Consistency of DFFS-$\Lambda$CDM and CFS-$\Lambda$CDM}
\label{sec:inside}
\begin{figure}[t]
\caption{\label{fig:puffers}Cosmological quantities in the CFS,
  relative to the DFFS, if the stellar population contributes with
  $\mathcal{P}_s \equiv w_s \rho_s$.  Pressures $\mathcal{P}_s
  \lesssim 10^{-4}\rho_s$ are conservative (i.e. very large) for
  gravitationally bound systems at late times.  Stellar growth follows
  \citet{MadauDickinson2014}.  The palette gives the equation of state
  parameter $w_s$.  As can be seen, the CFS is observationally
  indistinguishable from $\Lambda$CDM.  (Top panel) Difference in
  comoving energy density relative to the DFFS.  The dotted horizontal
  line gives an observational upper bound for present day values.
  (Bottom panel) Fractional difference of expansion history relative
  to the DFFS.  The dotted horizontal line gives an observational
  upper bound for present day values.}
\resizebox{\linewidth}{!}{\includegraphics{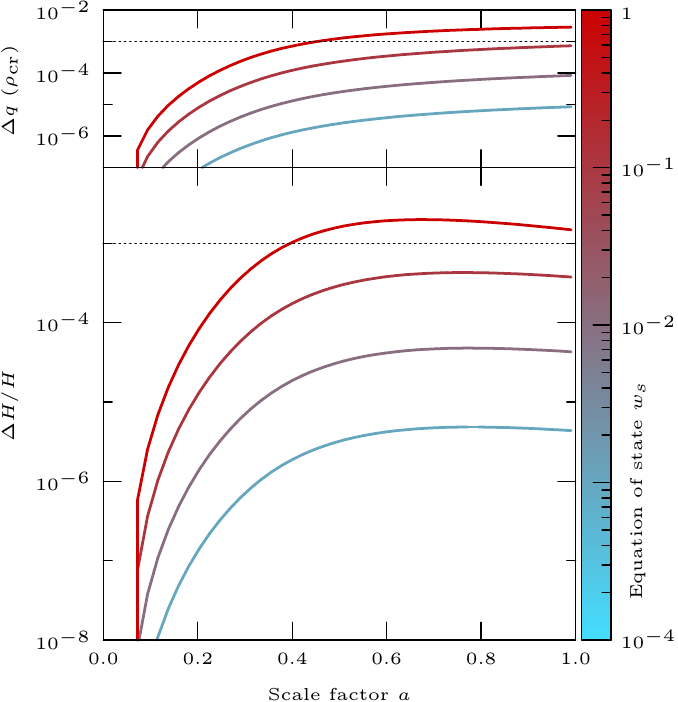}}
\end{figure}
In this Appendix, we demonstrate that $\Lambda$CDM with the CFS
remains observationally viable.  As established in
\S\ref{sec:darkages} and discussed in Appendix~\ref{sec:assumptions},
there is no need to track pressures until the non-linear growth
regime.  Before this regime, therefore, the DFFS and the CFS are
equivalent.  Once into the non-linear regime, predictions are not made
with linear perturbation theory.  Instead, $N$-body simulations with
initial conditions determined by linear perturbation theory are
required.  We thus show that the standard Newtonian behaviour on an
expanding background is observationally unchanged.  To do this, we
show that the ``Hubble drag'' and density contrast are observationally
equivalent in the DFFS and the CFS.

Our approach in this section will be to conservatively bound the
influence of pressures from typical systems on both cosmological and
peculiar motions.  We will quantify the influence on cosmological
motion with the fractional difference in the Hubble rate
\begin{align}
\frac{\Delta H}{H} \equiv \frac{H_{\Lambda\mathrm{CDM}} - H_\mathrm{test}}{H_{\Lambda\mathrm{CDM}}}\label{eqn:DeltaH}.
\end{align}
We will quantify the influence on peculiar motions with the difference
in total comoving energy densities
\begin{align}
\Delta q \equiv q_{\Lambda\mathrm{CDM}} - q_\mathrm{test}.
\end{align}
Here $q$ is defined in Eqn.~(\ref{eqn:q}) as the total density
appearing in the Friedmann energy Eqn.~(\ref{eqn:fried1}).  We will
use DFFS-$\Lambda$CDM (with purely collisionless matter) as the
reference expansion, and consider only Planck best-fit cosmological
parameters for $\Omega_m$ and $\Omega_\Lambda$~\citep{Planck2015XIII}.

To build an extremely conservative upper bound on the effect of
pressures neglected in the DFFS, we consider the densest stars, the
neutron stars.  While the neutron star equation of state is an active
area of research, example stable polytrope
models~\citep[e.g.][]{hansen2012stellar} give a local core pressure of
$\bar{\mathcal{P}} \sim 10^{-6}\bar{\rho}$.  For simplicity, we will
regard the equation of state parameter $w_s$ as constant during the
following analysis.

We define a depletion fraction by dividing the cosmic stellar density,
given in derivative form in Eqn.~(\ref{eqn:dStardz}), by $\Omega_m$.
We then numerically integrate Eqn.~(\ref{eqn:dynamical}).  We choose
extremely conservative (i.e. large) values for the equation of state
parameter
\begin{align}
10^{-4} < w_s < 1.
\end{align}
The results for the expansion history are shown in the bottom panel of
Figure~\ref{fig:puffers}.  It is clear that keeping positive pressures
in the CFS cannot observably alter cosmological motions.

From Eqn.~(\ref{eqn:DeltaH}) and Figure~\ref{fig:puffers}, the expansion
loses energy relative to the DFFS with fixed comoving densities.  From
conservation of energy and momentum given in Eqn.~(\ref{eqn:cons}), 
\begin{align}
\derp{\rho_s} + \derp{\rho_0} = -3Hw_s\rho_s,
\end{align}
it is clear that $\rho_s$ can only remain constant at fixed $w_s$ when
$\rmd\rho_0/\rmd t \propto H$.  Therefore, a small amount of energy
density $\Delta q$ is subsequently lost after the conversion of
collisionless stress $\rho_0$ to stress under compression $\rho_s$.
This loss, relative to the DFFS, is shown in the top panel of
Fig.~\ref{fig:puffers}.  To understand the influence of this loss,
recall that peculiar motions are sourced by the energy density
contrast~\citep[\S7]{peebles1980large}
\begin{align}
\nabla^2 \phi = \frac{3}{2}\left[\bar{\rho}(\mathbf{x}, a) - \frac{q(a)}{a^3}\right]a^2.
\end{align}
Here $\bar{\rho}(\mathbf{x}, a)$ is the total local density, $\phi$ is the
Newtonian gravitational potential, and $q$ is the (background)
comoving energy density defined in Eqn.~(\ref{eqn:q}).  In bound
systems at late times,
\begin{align}
\bar{\rho}(\mathbf{x}, a) \ggg \frac{q(a)}{a^3} \gg \frac{\Delta q(a)}{a^3}.
\end{align}
It is immediately clear that $\Delta q$ can have no observable effect
on the local dynamics of any contemporary system.

In summary, we have demonstrated that even conservative (i.e. large)
estimates of pressure contributions to the CFS produce no observable
discrepancies from the pressure-free DFFS.  This is true both for
cosmological motions, as characterized by $H$, and for peculiar
motions as determined by the density contrast.  These results
establish that the CFS is observationally consistent with the DFFS
under the assumptions of $\Lambda$CDM.

\subsection{Friedmann's equations}
\label{sec:friedmann}
Here we derive the form of Friedmann's equations used in our numerical
analysis.  Units will remain explicit until
Eqn.~(\ref{eqn:dynamical}), which is written with the units used
elsewhere in this paper.  We begin with the spatially flat
Robertson-Walker line element
\begin{align}
g_{\mu\nu} = -\rmd{t}^2 + a(t)^2\rmd\mathbf{x}^2
\label{eqn:rw}
\end{align}
with $a(t)$ dimensionless.  Substitution of
Eqns.~(\ref{eqn:star_source}), (\ref{eqn:normal_source}), and
(\ref{eqn:rw}) into Einstein's equations with cosmological constant
term $\Lambda$ gives the Friedmann equations
\begin{align}
H^2 &= \frac{\gamma^2}{a^3}\left[\rho_0(a) + \rho_s(a) + \frac{\Lambda}{3\gamma^2} a^3\right] \label{eqn:fried1}\\
H^2 + 2\frac{\ddot{a}}{a} &= -\frac{3\gamma^2}{a^3}\left[ \mathcal{P}_s(a) - \frac{\Lambda}{3\gamma^2} a^3\right] \label{eqn:fried2}
\end{align}
where dot denotes derivative with respect to coordinate time.  Here $H
= \dot{a}/a$ and $\gamma^2 \equiv 8\pi G/3$.  Define the total
comoving energy density $q$, effective equation of state
$w_s$, and scale velocity $v$ by
\begin{align}
q &\equiv \rho_0(a) + \rho_s(a) + \frac{\Lambda}{3\gamma^2} a^3 \label{eqn:q}\\ 
P_s &\equiv w_s\rho_s(a) \label{eqn:w_s}\\
v &\equiv \derp{a}. \label{eqn:v}
\end{align}
Substitution of Eqns.~(\ref{eqn:q}) and (\ref{eqn:v}) into
Eqn.~(\ref{eqn:fried1}) gives an algebraic constraint
\begin{align}
v^2 &= \frac{q}{a}. \label{eqn:algebraic}
\end{align}
Substitution of Eqns.~(\ref{eqn:rho0}), (\ref{eqn:w_s}), and
(\ref{eqn:algebraic}) into Eqn.~(\ref{eqn:fried2}) gives a single
dynamical equation, which we express in terms of $a$
\begin{align}
\frac{\rmd v}{\rmd a} &= \frac{3a\Omega_\Lambda}{2v}(1+w_s) + \frac{3w_s\Omega_m}{2va^2}(1-\Delta(a)) - \frac{v}{2a}(3w_s + 1). \label{eqn:dynamical}
\end{align}
The dynamics described by Eqn.~(\ref{eqn:dynamical}) are completely
general.

\section{Causality within the CFS}
\label{sec:causality}
In this Appendix, we discuss causality in the perturbative expansion
about a RW background, given the CFS defined in
Appendix~\ref{sec:action}.  GR possesses a well-posed initial value
formulation \citep[e.g.][\S10]{WaldGR}.  In other words, starting from
reasonable initial data, the metric and stress-energy evolve uniquely
and causally.  For regular (non-singular)
perturbations \citep[e.g.][\S7.1]{lin1988mathematics}, one expands the
relevant dynamical quantities in a perturbation series
\begin{align}
g_{\mu\nu} &\equiv \sum_{n=0}^\infty \xo{(n)}{g_{\mu\nu}} \\
T_{\mu\nu} &\equiv \sum_{n=0}^\infty \xo{(n)}{T_{\mu\nu}}.
\end{align}
Since the dynamical equations of GR are non-linear, it is not true
that each term in the expansion must evolve causally.  Only the
converged series are subject to this constraint.

Consider the formation of a single gravastar, described in a
perturbation series with the $(0)$-order metric defined to be RW.  As
detailed in Appendix~\ref{sec:action}, the transition of baryonic
matter into vacuum will change the $(0)$-order source.  This change
is, by definition, acausal since the $(0)$-order source contains no
position dependence.  By construction of the first-order theory, the
linearized $(1)$-order perturbations (on top of any background)
propagate causally.  If there are acausal features in the $(0)$-order
solution, it must be that the higher order perturbations act to
restore the causal character of the entire converged series.  We do
not attempt a general proof of this statement, but use the DE problem,
and previous attempts at its resolution within GR, to provide
plausibility for the above assertion.

\citet{buchert2012backreaction} review arguments that the higher-order
terms in the perturbative treatment mimic DE through ``backreation.''
\citet{ishibashi2005can}, however, argued years earlier that this
effect is far too small.  What pique's our attention, however, is that
\citet[][\S3]{rasanen2004dark} found that the effect is of the wrong
sign.  In fact, this is consistent with the gauge invariant findings
of \citet{abramo1997energy}, who report a negative backreaction at
primordial times.

Given our developed gravastar context, we speculate that the role of
backreation is to transiently cancel the imprint of newly formed
gravastars from the $(0)$-order source.  Note that this effect is
localized in spacetime.  This is because stars are collapsing to
gravastars everywhere, and so there is a mean separation between
progenitors that is small compared to the horizon.

\section{BH population estimation from the stellar population}
\label{sec:distribution}
\input{distribution.tex}

\section{Systems with large velocity dispersion}
\label{sec:dispersion}
We wish to compute an order of magnitude pressure contribution from
large, virialized systems, like rich clusters.  Pressure supported
clusters are regarded as ideal gases
\begin{equation}
\mathcal{P}_c = nkT
\end{equation}
where $\mathcal{P}_c$ is the pressure, $n$ is the number density, $k$ is
Boltzmann's constant, and $T$ is the temperature.  Let $m$ be a
typical mass, then
\begin{equation}
\mathcal{P}_c = \frac{\rho_c kT}{m}.
\end{equation}
Assuming a monoatomic gas, the average energy is
\begin{equation}
E = \frac{3}{2} kT \label{eqn:avgE}.
\end{equation}
One can construct a Newtonian kinetic energy from the velocity
dispersion $\sigma$
\begin{equation}
E = \frac{1}{2}m\sigma^2 \label{eqn:disp}.
\end{equation}
Combining Eqns.~(\ref{eqn:avgE}) and (\ref{eqn:disp}) gives the
temperature
\begin{equation}
kT = \frac{m\sigma^2}{3}.
\end{equation}
To determine a reasonable upper bound, rich clusters can have $\sigma
\sim 10^3~\mathrm{km}/\mathrm{s}$~\citep{struble1999compilation}.
Since we work in natural units, $\sigma^2 \to (\sigma/c)^2 \sim
10^{-5}$.  Thus, the kinetic pressure of a rich cluster goes as
\begin{equation}
\mathcal{P}_c \lesssim \frac{10^{-5}}{3}\rho_c.
\end{equation}

\bibliographystyle{aasjournal}
\bibliography{oprcp3}
\end{document}

%% file: distribution.tex
In this Appendix, we develop a relation between the stellar population
and the BH population.  This relation is not required to
experimentally test the predictions of this paper.  The only relevant
observable $\Delta_\mathrm{BH}$ will soon be constrained directly
through gravitational wave astronomy.  Relation to the stellar
population, however, is natural since BHs come from stars.  Further, a
quantitative model will allow us to check consistency over a diverse
set of existing astrophysical data.  We will formally outline the
procedure, and then construct a usable model.

Denote the comoving coordinate number density of galaxies with mass in
the range $\rmd M_g$ by
\begin{align}
\frac{\rmd n_g}{\rmd M_g}~\rmd M_g \label{eqn:ngMg}
\end{align}
Denote the \emph{number count} of BHs with mass in the range $\rmd
M_\mathrm{BH}$ in a galaxy of mass $M_g$ as
\begin{align}
\frac{\rmd N_\mathrm{BH}(M_g, M_\mathrm{BH})}{\rmd M_\mathrm{BH}}~\rmd M_\mathrm{BH}. \label{eqn:BHs}
\end{align}
We re-express this relation in terms of stars.  Following
\citet{fryer2001theoretical}, we use the fact that BHs are sampled from the same distribution as stars to write
\begin{align}
\frac{\rmd N_\mathrm{BH}}{\rmd M_\mathrm{BH}} = \frac{\rmd N_*}{\rmd M_*}\frac{\rmd M_*}{\rmd M_\mathrm{BH}}. \label{eqn:kalogera}
\end{align}
Assembling Eqns.~(\ref{eqn:ngMg}), (\ref{eqn:BHs}), and
(\ref{eqn:kalogera}), the number count of BHs with mass in the range
$\rmd M_\mathrm{BH}$ across galaxies in the mass range $\rmd M_g$ is
\begin{align}
\frac{\rmd n_g}{\rmd M_g} \frac{\rmd N_*}{\rmd M_*}(M_g, M_*) \frac{\rmd M_*}{\rmd M_\mathrm{BH}}~\rmd M_\mathrm{BH}\rmd M_g.
\end{align}
To form a population average comoving coordinate density over some
observable $X(M_\mathrm{BH})$, we multiply by the observable and
integrate over all $BH$ masses and galaxy masses
\begin{align}
\left<X\right> \equiv \int_{m_g}^\infty \frac{\rmd n_g}{\rmd M_g} \int_{m_\mathrm{BH}}^\infty X(M_\mathrm{BH}) \frac{\rmd N_*}{\rmd M_*}(M_g, M_*) \frac{\rmd M_*}{\rmd M_\mathrm{BH}}~\rmd M_\mathrm{BH}\rmd M_g. \label{eqn:generic_population_average}
\end{align}
Note that accretion effects do not change the number of objects, but
that mergers do.  Mergers also affect the total mass through radiative
losses, which can be around $5\%$ of the final
remnant \citep[e.g.][]{abbott2016observation}.  In the interest of
simplicity, we do not attempt to take account of these effects.

\begin{deluxetable}{lccc}
\tablewidth{0pt}  
\tabletypesize{\footnotesize}
\tablecaption{\label{tbl:distribution_parameters}Astrophysical parameters for BH population estimation}
\tablehead{\colhead{Parameter} & \colhead{Value} & \colhead{Units} & \colhead{Reference}}
\startdata
$\alpha$ & -2.35 & - & \citet{salpeter1955luminosity} \\
$R$ & 0.27 & - & \citet{MadauDickinson2014} \\
$m_c$ & $0.1$ & $M_\odot$ & \citet{chabrier2003galactic} \\
$m_\mathrm{TOV}$ & $3$ & $M_\odot$ & \citet{kalogera1996maximum} \\
\enddata
\end{deluxetable}

To produce a quantitatively useful model from
Eqn.~(\ref{eqn:generic_population_average}), we note that
\begin{align}
\int_{m_c}^\infty \frac{\rmd N_*(M_g, M_*)}{\rmd M_*} M_* ~\rmd M_* \equiv (1-R)M_g \label{eqn:galaxy_mass}
\end{align}
where $R$ is the fraction of stellar mass returned to the host galaxy
upon star death.  Note that $m_c$ is a cutoff mass for the stellar
distribution.  We next assume a power-law for this distribution
\begin{align}
\frac{\rmd N_*(M_g, M_*)}{\rmd M_*} \equiv B M_*^\alpha
\end{align}
with the normalization $B$ determined by Eqn.~(\ref{eqn:galaxy_mass}).
The result is
\begin{align}
\frac{\rmd N_*(M_g, M_*)}{\rmd M_*} = -(2+\alpha)\frac{M_g(1-R)}{m_c^{\alpha + 2}} M_*^\alpha. \label{eqn:*_distribution}
\end{align}
Since \citet{chabrier2003galactic} finds a flattening of the stellar IMF around
$0.1M_\odot$, we will take 
\begin{align}
m_c \equiv 0.1M_\odot
\end{align}
to maintain the validity of our power-law assumption.  Reasonable
values for $R$ and $\alpha$ are given in
Table~\ref{tbl:distribution_parameters}.  Inserting
Eqn.~(\ref{eqn:*_distribution}) into
Eqn.~(\ref{eqn:generic_population_average}) and integrating over the
host galaxy masses, we find
\begin{align}
\left<X\right> = -\rho_g(a)(1-R)\frac{(2+\alpha)}{m_c^{\alpha + 2}} \int_{m_\mathrm{BH}}^\infty X(M_\mathrm{BH}) M_*^\alpha \frac{\rmd M_*}{\rmd M_\mathrm{BH}}~\rmd M_\mathrm{BH} \label{eqn:useable_form}
\end{align}
where $\rho_g(a)$ is the comoving stellar
density \citep[e.g.][Fig.~11]{MadauDickinson2014}.  To proceed further,
we require $M_\mathrm{BH}(M_*)$.  This has been recently estimated
through stellar collapse simulations by
\citet[][Fig.~4]{fryer2012compact}.  We consider a linear ansatz for
simplicity
\begin{align}
M_\mathrm{BH}(M_*) \equiv kM_* - |b| \label{eqn:MBH_ansatz}
\end{align}
and give the fit parameters in Table~\ref{tbl:fryer_fits}.  For ansatz
Eqn.~(\ref{eqn:MBH_ansatz}), $b < 0$ and $k > 0$ are the only
physically allowed values.

\begin{deluxetable}{llccc}
\tablewidth{0pt}  
\tabletypesize{\footnotesize}
\tablecaption{\label{tbl:fryer_fits}Linear fits to $M_\mathrm{BH}(M_*)$}
\tablehead{\colhead{Metallicity} & \colhead{Model} & \colhead{Slope $k$} & \colhead{$y$-Intercept $|b|$ ($M_\odot$)} & \colhead{$\chi_\nu^2$}}
\startdata
$Z = Z_\odot$ & rapid & $0.19 \pm 0.06$ & $0.038 \pm 1.8$ & $9.2$ \\
 & delayed & $0.23 \pm 0.03$ & $1.1 \pm 0.96$ & $2.3$ \\
$Z = 0$ & rapid & $1.2 \pm 0.15$ & $18 \pm 3.7$ & $34$ \\
 & delayed & $1.2 \pm 0.22$ & $18 \pm 5.3$ & $70$ \\
\enddata
\tablecomments{parameters characterize a coarse fit $M_\mathrm{BH}=k M_* - |b|$ to $M_\mathrm{BH}(M_*)$ based on the stellar collapse simulations of \citet{fryer2001theoretical}}
\end{deluxetable}

The most important case for our concerns is
\begin{align}
X(M_\mathrm{BH}) \equiv M_\mathrm{BH}^q \qquad q \in \mathbb{Q}.
\end{align}
For example, $q=1$ will estimate the cosmological
$\rho_\mathrm{BH}$.  Substituting Eqn.~(\ref{eqn:MBH_ansatz}) into
Eqn.~(\ref{eqn:useable_form}), we find
\begin{align}
\left<M_\mathrm{BH}^q\right> = -\rho_g(a)(1-R)\frac{(2+\alpha)}{m_c^{\alpha + 2}} \int_{m_*}^\infty (kM_* - |b|)^q M_*^\alpha~\rmd M_* \label{eqn:dimful_average}
\end{align}
where we have switched to integration over the progenitors.  In the
relevant simulations of \citet{fryer2012compact}, the lower bound for
the progenitors is $11M_\odot$.  For our purposes, we will demand that
the remnant object exceed the Tolman-Oppenheimer-Volkoff limit
$m_\mathrm{TOV}$ for neutron-degeneracy pressure supported systems.
Inspection of \citet[][Fig.~4]{fryer2012compact} gives the following
reasonable cutoffs
\begin{align} m_* =
\begin{cases}
11M_\odot & Z=Z_\odot \\
16M_\odot & Z=0 
\end{cases}\label{eqn:cutoffs}
\end{align}
where $Z$ is the stellar metallicity.  Note that
\begin{align}
m_0 \equiv |b|/k
\end{align}
gives the natural cutoff for our linear ansatz.  This cutoff is
aphysical because it corresponds to a massless remnant, but is
theoretically useful to estimate systematics from
Eqn.~(\ref{eqn:cutoffs}). 

\subsection{Determining $k$, $|b|$, and $m_*$ as functions of redshift}

\begin{deluxetable}{llcc}
\tablewidth{0pt}  
\tabletypesize{\footnotesize}
\tablecaption{\label{tbl:time-evol}Metallicity dependence of remnant population parameters}
\tablehead{\colhead{Parameter} & \colhead{Model} & \colhead{Normalization $B$} & \colhead{$z$-constant $\zeta$}}
\startdata
$k$ & rapid & $0.19$ & $0.09$ \\ & delayed & $0.23$ & $0.083$ \\ $|b|$
& rapid & $0.038$ & $0.31$ \\ & delayed & $1.11$ & $0.14$ \\ $m_*$&
Eqn.~(\ref{eqn:cutoffs}) & $11$ & $0.01$ \\
\enddata
\tablecomments{Fits are exponential $B\exp(\zeta z)$, assuming $Z(z = 20) \equiv 0$ (i.e. neglecting primordial metal abundances)}
\end{deluxetable}

Since \citet{fryer2012compact} only provide two metallicities,
unfortunately we must interpolate the time-evolution of $k$, $|b|$,
and $m_*$ between only two points in time.
From \citet[][Fig.~14]{MadauDickinson2014}, $Z(z)$ is reasonably
well-approximated by an exponential.  The simplest assumption is that
$k$, $|b|$, and $m_*$ are also exponentials
\begin{align}
\left\{k, |b|, m_*\right\}(z) \equiv \{B\} \exp\left(\{\zeta\} z\right).
\end{align}
Here $\{B\}$ are the normalizations and $\{\zeta\}$ are redshift ``time
constants.''  We present these fit parameters in
Table~\ref{tbl:time-evol} to $k$, $|b|$, and $m_*$ for both rapid and
delayed supernovae models.

A more rigorous analysis would consider a range of monotonic
interpolations to determine the systematic error introduced in all
subsequent quantities involving $k$, $|b|$, and $m_*$. We forgo this
procedure since a single additional
$M_\mathrm{rem}(M_\mathrm{prog})(Z)$ for $Z \in (0, Z_\odot)$ would
mitigate this necessity.

\subsection{Cosmological BH density}
\label{sec:proxy_BH_density}

\begin{deluxetable}{llccc}
\tablewidth{0pt}  
\tabletypesize{\footnotesize}
\tablecaption{\label{tbl:collapse_fractions}Fraction of comoving stellar
  density collapsed into BH, based on metallicity and
  supernovae engine}
\tablehead{\colhead{Metallicity} & \colhead{Model} & \colhead{$m_0~(M_\odot)$} & \colhead{$\Xi(m_*)~(\%)$} & \colhead{$\Xi(m_0)~(\%)$}}
\startdata
$Z = Z_\odot$ & rapid & $0.15$ & $2.6$ & $4~(m_\mathrm{TOV})$\\
 & delayed & $4.8$ & $2.8$ & $3.2$ \\
$Z = 0$ & rapid & $15.7$ & $11$ & $11$\\
 & delayed & $15.2$ & $12$ & $11$ \\
\enddata
\tablecomments{Astrophysical parameters used are given in
  Table~\ref{tbl:distribution_parameters}.  For comparison,
  \citet{BetheBrown1994} predict a late-time ($Z \sim Z_\odot$) collapse
  fraction $\Xi \sim 1\%$.}
\end{deluxetable}

The $q=1$ case corresponds to the cosmological comoving coordinate BH
mass density.  Usefully, $q=1$ can be integrated by hand in closed
form
\begin{align}
\rho_\mathrm{BH}(a) &= (1-R)\left[k\left(\frac{m_*}{m_c}\right)^{\alpha + 2} - \frac{|b|}{m_c}\left(\frac{\alpha + 2}{\alpha + 1}\right)\left(\frac{m_*}{m_c}\right)^{\alpha + 1}\right]\rho_g(a) \label{eqn:stellar_bh_model}\\
&\equiv \Xi(a)\rho_g(a) \label{eqn:Xi_defn}
\end{align}
where we have defined the collapse fraction $\Xi(a)$.  Time variation
in $\Xi$ can come from $\alpha$, $k$, $|b|$, and $m_*$.  For the
collapse models detailed in Table~\ref{tbl:fryer_fits}, we give cutoff
parameters and collapse fractions in
Table~\ref{tbl:collapse_fractions}, based on the astrophysical
parameters given in Table~\ref{tbl:distribution_parameters}.  We make
no effort to propagate errors due to the coarseness of the fit.  

To quantify the sensitivity of these data to the cutoffs given in
Eqn.~(\ref{eqn:cutoffs}), note that the integrand remains finite for
$q=1$.  Thus, we may naturally remove the cutoff and set $m_* \equiv
m_0$.  We have done this except where $m_0 < m_\mathrm{TOV}$, where we
take $m_* \equiv m_\mathrm{TOV}$.  In all cases, the solar metallicity
results are consistent with $\Xi \sim 1\%$ as computed
by \citet{BetheBrown1994} via entirely different means.

We have assumed zero time lag between the stellar population and the
BH population.  Let us justify this simplification for $0.2 \ll z <
20$.  This places us within matter domination, so we may approximate
\begin{align}
H \simeq \frac{\sqrt{\Omega_m}}{a^{3/2}}.
\end{align}
Considering the differentials, we see that
\begin{align}
\rmd a = \frac{\sqrt{\Omega_m}}{a^{1/2}} \rmd t
\end{align}
and so we may approximate the lag in scale $a_L$ relative to the lag
in time $\tau_L$ as
\begin{align}
a_L \simeq \frac{\sqrt{\Omega_m}}{a^{1/2}} \tau_L.
\end{align}
According to dimensional arguments, for stars massive enough to
collapse to BH, $\tau_L \lesssim 10^7$ years at metallicity $Z_\odot$.
Given our choice in time unit, this becomes the upper bound
\begin{align}
\tau_L \lesssim \frac{6 \times 10^{14}~\mathrm{s}}{5 \times
  10^{17}~\mathrm{s}} \sim 10^{-3}.
\end{align}
Note that this is a conservative upper bound on primordial star
lifetimes.  To justify the neglect of the time lag, we must have that
\begin{align}
\rho_g(a - a_L(a)) = \rho_g[a(1 - a_L/a)] \simeq \rho_g(a)
\end{align}
and so we require that $a_L/a \ll 1$.  A sufficient condition is then
\begin{align}
\frac{a_L}{a} < \frac{\sqrt{\Omega_m}}{a^{3/2}} 10^{-3} \ll 1, 
\end{align}  
which implies
\begin{align}
z &\ll 148.
\end{align}
For representative numbers, at $z = 20$ the time lag is a $5\%$
correction, but at $z = 5$ the time lag is a $0.8\%$ correction.  Most
importantly, at the peak of star formation near $z\sim 2$, the time
lag is a $0.2\%$ correction.  We may thus consider
\begin{align}
\rho_{BH}(a) = \Xi(a)\rho_g(a) \qquad (z < 5) \label{eqn:BHvsStar}
\end{align}
to good precision.  

\begin{deluxetable}{lccc}
\tablewidth{0pt}  
\tabletypesize{\footnotesize}
\tablecaption{\label{tbl:units}Adopted astrophysical parameters}
\tablehead{\colhead{Quantity} & \colhead{Value} & \colhead{Units} & \colhead{Reference}}
\startdata
$G$ & $6.67 \times 10^{-11}$ & $\mathrm{m}^3~\mathrm{kg}^{-1}~\mathrm{s}^{-2}$ & \citet{patrignani2016review} \\
$c$ & $3.00 \times 10^{8}$ & $\mathrm{m}~\mathrm{s}^{-1}$ & \citet{patrignani2016review} \\
$M_\odot$ & $1.99 \times 10^{30}$ & $\mathrm{kg}$ & \citet{patrignani2016review} \\
$H_0$ & $69.3$ & $\mathrm{km}~\mathrm{s}^{-1}~\mathrm{Mpc}^{-1}$ & \citet{Planck2015XIII} \\
$M_\odot \mathrm{Mpc}^{-3} \mathrm{yr}^{-1}$ & $0.277$ & $\rho_\mathrm{cr} H_0$ & - \\
\enddata
\tablecomments{Values used to construct unit conversions between $H_0 = \rho_\mathrm{cr} \equiv 1$ and common literature units.   }
\end{deluxetable}

In practice, we will be interested in the late-time behavior of these
quantities.  Since $\rho_\mathrm{BH}$ depends on the entire conversion
history, neglecting accretion could have a substantial effect on
$\rho_\mathrm{BH}$.  Since \citet{LiAccretion2006} find that accretion
effects diminish significantly below $z < 5$, it is more convenient to
examine $\rmd \rho_\mathrm{BH}/\rmd a$
\begin{align}
\frac{\rmd \rho_\mathrm{BH}}{\rmd a} = -\Xi(a)\frac{\rmd \rho_g}{\rmd z}\frac{1}{a^2} + \frac{\rmd \Xi}{\rmd a}\rho_g.
\end{align}
A global fit, performed by \citet{MadauDickinson2014}, to very many
stellar surveys gives the comoving stellar density as
\begin{align}
\frac{\rmd \rho_g}{\rmd z} &= 0.277\frac{(R - 1)\psi}{(1+z)H} \label{eqn:dStardz}\\
\psi &\equiv 0.015 \frac{(1+z)^{2.7}}{1 + [(1+z)/2.9]^{5.6}} \\
R &\equiv 0.27.
\end{align}
The factor of $0.277$ converts units as listed in
Table~\ref{tbl:units}.  We determine $H$ given the Planck assumed dark
energy equation of state Eqn.~(\ref{eqn:planck_w})
\begin{align}
H(z) &\equiv \sqrt{\Omega_m(1+z)^3 + \Omega_\Lambda(1+z)^{3[1+w_\mathrm{eff}(0)]}\exp\left(-\frac{3w_az}{z+1}\right)}.
\end{align}

\subsection{Expectation values with generic $q$}
It is useful to define notation to compress and de-dimensionalize
\begin{align}
\lambda(q) &= q + \alpha + 1 \\
r &\equiv m_0/m_*.
\end{align}
We may now write the de-dimensionalized Eqn.~(\ref{eqn:dimful_average}) 
\begin{align}
\left<M_\mathrm{BH}^q\right> &= \rho_g(a)(R-1)\frac{(2+\alpha)}{m_c^{\alpha + 2}}\frac{|b|^{\lambda(q)}}{k^{\alpha+1}}\int_{1/r}^{\infty}(z-1)^qz^\alpha~\rmd z. \label{eqn:dedim_average}
\end{align}
Notice that $z \geq 1$ always, and so the integrand remains real for
all $q,\alpha \in \mathbb{R}$.  We then perform an inversion $z \to 1/x$ to make the
domain of integration finite
\begin{align}
\left<M_\mathrm{BH}^q\right> = \rho_g(a)(R-1)\frac{(2+\alpha)}{m_c^{\alpha + 2}}\frac{|b|^{\lambda(q)}}{k^{\alpha+1}} \int_0^r (1-x)^qx^{-\lambda(q)-1}~\rmd x.
\end{align}
Define the following notation
\begin{align}
_2F_1\left\{q\right\} &\equiv {_2F_1}\left(-q, -\lambda(q); -(q + \alpha); r\right).
\end{align}
We may then write
\begin{align}
\int_0^r (1-x)^qx^{-\lambda(q)-1}~\rmd x &=-\frac{{_2F_1}\{q\}}{\lambda(q)}r^{-\lambda(q)} 
\end{align}
and finally we find 
\begin{align}
\left<M_\mathrm{BH}^q\right> &= \rho_g(a)(1-R)\frac{(2+\alpha)m_*^{\lambda(q)}k^q}{m_c^{\alpha + 2}} \frac{{_2F_1}\left\{q\right\}}{\lambda(q)}.
\end{align}
Values for $_2F_1\left\{q\right\}$ can be computed in any modern CAS.